\documentclass[preprint]{aastex}
\def\day{{}$^{\rm d}$\llap{.}}
\def\lsun{{L_\odot}}
\def\msun{{M_\odot}}
\begin{document}

\title{THE MACHO PROJECT LARGE MAGELLANIC CLOUD VARIABLE STAR INVENTORY. 
XIII. FOURIER PARAMETERS FOR THE FIRST OVERTONE RR LYRAE VARIABLES AND 
THE LMC DISTANCE}

\author{
      C.~Alcock\altaffilmark{1,2},
    D.R.~Alves\altaffilmark{3},
    T.S.~Axelrod\altaffilmark{4},
    A.C.~Becker\altaffilmark{5},
    D.P.~Bennett\altaffilmark{2,6},
    C.M.~Clement\altaffilmark{7},
    K.H.~Cook\altaffilmark{2},
    A.J.~Drake\altaffilmark{2,8},
    K.C.~Freeman\altaffilmark{9},
      M.~Geha\altaffilmark{2,10},
      K.~Griest\altaffilmark{2,11},
    M.J.~Lehner\altaffilmark{1},
    S.L.~Marshall\altaffilmark{2},
      D.~Minniti\altaffilmark{8},
    A.~Muzzin\altaffilmark{7},
    C.A.~Nelson\altaffilmark{2},
    B.A.~Peterson\altaffilmark{9},
      P.~Popowski\altaffilmark{12},
    P.J.~Quinn\altaffilmark{13},
    A. W. Rodgers\altaffilmark{14},
    J.F.~Rowe\altaffilmark{15},
      W.~Sutherland\altaffilmark{16},
      T.~Vandehei\altaffilmark{11},
    D.L.~Welch\altaffilmark{17}
        }

%-------------------- institutions and email ----------------------
\altaffiltext{1}{Department of Physics and Astronomy, University of
Pennsylvania, PA 19104\\}

\altaffiltext{2}{Lawrence Livermore National Laboratory, Livermore, CA 94550\\}

\altaffiltext{3}{Columbia Astrophysics Laboratory, Columbia University, 
550 West 120th Street, Mailcode 5247, New York, NY 10027\\
    Email: {\tt alves@astro.columbia.edu}}

\altaffiltext{4}{Steward Observatory, University of Arizona, Tucson, AZ
85721 \\}

\altaffiltext{5}{Bell Laboratories, Lucent Technologies, 600 Mountain Avenue,
Murray Hill, NJ 07974\\}

\altaffiltext{6}{Department of Physics, University of Notre Dame, 
Notre Dame, IN 46556\\}

\altaffiltext{7}{Department of Astronomy \& Astrophysics, 
University of Toronto, Toronto ON, M5S 3H8, Canada\\
Email: {\tt cclement@astro.utoronto.ca, adam.muzzin@utoronto.ca}}

\altaffiltext{8}{Departmento de Astronomia, Pontifica Universidad Cat\'olica, 
Casilla 306, Santiago 22, Chile\\}

\altaffiltext{9}{Research School of Astronomy and Astrophysics, Australian
National University, Weston Creek, Canberra, ACT 2611, Australia\\}

\altaffiltext{10}{Department of Astronomy and Astrophysics, University of 
California, Santa Cruz, CA 95064\\}

\altaffiltext{11}{Department of Physics, University of California,
    San Diego, 9500 Gilman Drive, La Jolla, CA 92093-0354\\}

\altaffiltext{12}{Max-Planck-Institute for Astrophysics, Karl-Schwarschild-Str. 
1, Postfach 1317, 85741 G\"{a}rching, Germany \\}

\altaffiltext{13}{European Southern Observatory, Karl-Schwarzchild-Str.\ 2, 
        85748 G\"{a}rching, Germany\\}

\altaffiltext{14}{Deceased\\}

\altaffiltext{15}{Dept. of Physics \& Astronomy, University of British
Columbia, Vancouver BC, V6T 1Z4, Canada\\}

\altaffiltext{16}{Department of Physics, University of Oxford,
    Oxford OX1 3RH, U.K.\\}

\altaffiltext{17}{Department of Physics and Astronomy, McMaster University, 
Hamilton, Ontario Canada L8S 4M1\\}

\begin{abstract}
Shapes of RR Lyrae light curves can be described in terms of Fourier
coefficients which past research has linked with physical
characteristics such as luminosity, mass and temperature.
Fourier coefficents have been derived for the $V$ and $R$ light curves
of 785 overtone RR Lyrae variables in 16 MACHO fields near the bar
of the LMC. 
In general, the Fourier phase differences $\phi_{21}$, 
$\phi_{31}$ and $\phi_{41}$ increase and the amplitude ratio $R_{21}$
decreases with increasing period. The
coefficients for both the $V$ and $R$ magnitudes follow these
patterns, but the phase differences for the $R$ curves are on average
slightly greater, and their amplitudes are about 20\% smaller,
than the ones for the $V$ curves.
The $\phi_{31}$ and $R_{21}$ coefficients have been compared with
those of the first overtone RR~Lyrae variables in the  Galactic globular 
clusters NGC 6441, M107, M5, M3, M2, $\omega$~Centauri and M68.
The results indicate that many of the LMC variables have properties
similar to the ones in M2, M3, M5 and the Oosterhoff type I variables
in $\omega$~Cen, but they are different from the Oosterhoff type II
variables in $\omega$~Cen.
Equations derived from hydrodynamic pulsation models have been used to
calculate the luminosity and temperature for the 330
bona fide first-overtone variables. The results indicate that they
have $\log L$ in the range $1.6$ to $1.8\lsun$ and $\log T_{eff}$ 
between $3.85$ and $3.87$. Based on these temperatures, a
mean color excess $E(V-R) =0.08$ mag, equivalent to $E(B-V)=0.14$ mag, 
has been
estimated for these 330 stars. 
The 80 M5-like variables (selected according to their location in
the $\phi_{31}-\log P$ plot) are used to determine a LMC distance.
After correcting for the effects of extinction and 
crowding, a mean apparent magnitude $<V_0>=18.99 \pm 0.02$ (statistical)
$\pm 0.16$ (systematic) has been
estimated for these 80 stars. Combining this with
a mean absolute magnitude $M_V=0.56\pm 0.06$ for M5-like
stars derived from Baade-Wesselink analyses, main sequence fitting,
Fourier parameters and the trigonometric parallax of RR Lyrae,
we derive an LMC distance modulus $\mu=18.43\pm 0.06$ (statistical) $\pm
0.16$ (systematic) mag. The large systematic error arises from the difficulties 
of correcting for interstellar extinction and for crowding.
\end{abstract} 
\keywords{galaxies: distances and redshifts --- Magellanic clouds --- 
stars: variables:  RR Lyrae}

%
%************************ SECTION 1
%
\section{INTRODUCTION}

The MACHO Project database is a valuable
resource for studying the characteristics of variable stars
in the LMC. In paper II of this series, Alcock et al. (1996, hereafter
A96) identified 7900 RR Lyrae variables in twenty-two 
fields in the region of the LMC bar. 
The period-frequency distribution that they plotted for these variables
showed that the mode was 0\day 583, indicative
of an Oosterhoff (1939, 1944) type I population. In addition, there were two
other peaks in the distribution, at
0\day 342 and 0\day 281, which they attributed to  
variables pulsating in 
the first and second overtone modes, respectively.
The purpose of the present  investigation is to
perform a Fourier analysis of
the first-overtone (RR1\footnote{Throughout this
paper, we adopt the system of notation that
Alcock et al. (2000, hereafter A00) introduced for RR Lyrae
variables: RR0 for fundamental, RR1 for first-overtone, 
RR2 for second-overtone pulsators, etc.}) RR Lyrae
variables in order to determine an LMC distance. 
The LMC is a well known benchmark in the extragalactic
distance scale, and thus new measurements of its distance are
important in order to test the accuracy of standard cosmological models.

The distance to the LMC has a controversial history,
and yet in recent years a standard distance modulus has
emerged.  This is due in part to the 
completion of the {\it Hubble Space Telescope's}
key project to measure the Hubble constant with variable stars
and standard candles, which employs $\mu_{LMC}$ = 18.5 mag
(Freedman et al.~2001).
The Freedman et al.~(2001) result of $H_0 = 71 \pm 10$
km s$^{-1}$ Mpc$^{-1}$ (statistical and systematic 
error total) 
is in strikingly good agreement with that
derived from Wilkinson Microwave Anisotropy Probe data
($H_0 = 72 \pm 5$ km s$^{-1}$ Mpc$^{-1}$; Spergel et al.~2003).
These measurements of $H_0$
are based on entirely different physics, and thus their agreement
lends support to the accuracy of the standard LMC distance modulus
adopted by Freedman et al.~(2001).
It is, in fact, a recent trend in the literature
that most new LMC distance measurements
are in excellent agreement with the standard model, and in many cases
systematic errors in prior measurements are being found and corrected
(e.g., Alves et al.~2002, Mitchell et al.~2002).

In this investigation, we employ the Fourier decomposition technique, 
a method for quantifying the structural
characteristics of the observed light curves of variable stars.
It was first applied to 
RR Lyrae variables by Simon \& Teays (1982) who analysed the light
curves of 70 field RR Lyrae stars. 
Later, Clement, Jankulak \& Simon (1992) and Simon \& Clement 
(1993, hereafter SC93)
used the technique to compare the RR1 variables
in six Galactic globular clusters (GGCs) with metal abundances
ranging from [Fe/H]=$-0.99$ to $-2.17$ on the Zinn \&
West (1984, hereafter ZW) scale. In particular, they studied
the Fourier phase parameter $\phi_{31}$. 
By plotting $\phi_{31}$ versus $\log P$, 
they discovered that the clusters were segregated according to  
metallicity and that, within each cluster, 
$\phi_{31}$ increases with period. 
To understand the physical significance
of this result, SC93 analysed 
hydrodynamic pulsation models for first-overtone variables and found
that they could derive equations for expressing 
both the mass and the luminosity in terms of $\phi_{31}$ and
the pulsation period. An application of these equations to the
six GGCs indicated that there was a strong correlation between mean
RR1 luminosity and metal abundance of the cluster.
This provided independent evidence for the 
existence of an RR Lyrae luminosity-metallicity relation. It also demonstrated
that Fourier decomposition is a useful technique for estimating
the luminosity of an RR1 variable.

Most LMC distance determinations based on RR Lyrae variables depend on
the  luminosity-metallicity relation. In these studies, a mean metal abundance
must be adopted because
spectroscopic studies by A96, Bragaglia et al. (2001) and
Clementini et al. (2003, hereafter C03) have all shown that
there is a range of metal abundance among the field RR Lyraes in the LMC.
However,
since we do not have [Fe/H] values for the individual stars in our sample,
we take a different approach. In this investigation, our 
\it modus operandi \rm will be to compare
the Fourier parameters of the LMC RR1 variables
with the ones in some well-studied GGCs.
We will look for a subset of LMC RR1 variables that are similar to those
in one of these clusters.
Then we will assume that
their RR Lyrae variables have the same
mean absolute magnitude 
and use independent studies 
to determine the RR Lyrae absolute magnitudes. 

%
%************************ SECTION 2
%

\section{THE OBSERVATIONAL DATA}

Our investigation is based on the RR Lyrae data from 
16 LMC\footnote{The LMC
fields included are \#2, 3, 5, 6, 10, 11, 12, 13, 14, 15, 
18, 19, 47, 80, 81 and 82, all of which  are close to the bar.  
The field of view for each field is $0.52$ square degrees.
An identification chart and a
list of the R. A. and declination of the field centers 
are available at http://wwwmacho.mcmaster.ca}  
fields observed for the MACHO project. 
The program stars were selected from a preliminary sample that included all 
of the (approximately 1200) RR Lyrae variables deemed to be RR1 stars
according to their periods and light curve shapes.
These preliminary  data were instrumental magnitudes 
derived from the observations acquired between July 1992 and December 1995
through the MACHO $B_M$ and $R_M$ filters. 
The first step of our analysis was to derive the periods using
Stellingwerf's (1978)
phase dispersion minimization (PDM) technique and then to
fit both the $B_M$ and $R_M$  magnitudes to a Fourier series of the form:
\begin{eqnarray}
mag = A_0+\sum _{j=1}^{n} A_j \cos (j\omega t + \phi _j)  
                     \hskip 2mm 
\end{eqnarray}
where
$\omega$ is ($2\pi$/period), $t$ is the  time of the observation
and $n$ is the order of the fit. 
In each case, 
the order of the fit was 6 and all
magnitudes for which the assessed error was greater than $0.1$ mag
were excluded.
The phase differences, 
$\phi_{j1}= (\phi_j-j\phi_1)$ 
and amplitude ratios, $R_{j1}= (A_j/A_1$) 
were calculated and their standard errors 
were evaluated using the formulae of Petersen (1984). Since the
Fourier decomposition technique is not useful for studying stars with large
uncertainties in their coefficients, we included 
for further study only the stars with an 
error less than $0.3$ in $\phi_{21}$ or an error less than $0.4$ in 
$\phi_{31}$ in at least one of B or R. 
There were 785 stars that met these criteria. 
The instrumental magnitudes of these stars for
the observations obtained 
between July 1992 and December 1999 were transformed to the Kron-Cousins
V and R system using the equations derived by 
Alcock et al. (1999, hereafter A99). This calibration has been designated 
version 9903018.  
Calibration version numbers
may also appear in some MACHO database documentation 
and released light curve data 
(e.g., Allsman \& Axelrod 2001, Alcock et al.~2003).

%
%************************ SECTION 3 
%

\section{THE FOURIER ANALYSIS OF THE PROGRAM STARS}
\subsection{The Fourier Coefficients}

The calibrated $V$ and $R$ magnitudes for the 785 program stars 
extended over a longer time base than the preliminary data and so
we used the PDM technique to revise the periods before performing the
Fourier analysis.
Only observations obtained under good transparency
conditions were included. In addition,  all magnitudes
for which the assessed photometric error was greater than $0.1$ mag were
excluded.  The magnitudes were then fit to
equation (1) using a 6-order fit. 
It turned out that the sample included 105 stars that were found
to be multiperiodic\footnote{A00 introduced a new system of subclasses to
describe the frequency spectra of the multiperiodic variables: 
RR01 for double-mode 
(fundamental and first-overtone), RR12 for double-mode (first and 
second-overtone), RR1-PC for stars with period changes, RR1-BL for
Blazhko variables and 
RR1-$\nu1$, RR1-$\nu2$, RR1-$\nu$M for other multifrequency stars,
where 1, 2, M indicate that there are 1,  2 or more than 2
additional frequencies.}
by Kov\'acs and the MACHO collaboration (Kov\'acs et al. 2000, A00) 
so that only 680 stars in the sample were monoperiodic.
We present the data for these stars in Table 1.
For each star, we report the 
results of the Fourier analysis for both the $V$ and $R$ magnitudes. 
`N' denotes the number
of observations. The quantities $A_0$, $A_1$, $R_{j1}$,  $\phi_{j1}$ and
their standard
errors ($\sigma$), the amplitude  and $\sigma_{fit}$ were all obtained from the
fit of equation (1) to the data. 
The coefficients for the 105 multiperiodic
stars are listed in Tables 2 to 5, with the double-modes (RR01 and RR12) 
in Tables 2 and 3 respectively, the RR1-$\nu_1$ stars in Table 4 and other 
multifrequency variables in Table 5. 
For the rest of this investigation, we consider only the stars 
listed in Table 1.

Because there is overlap between some of the LMC fields in our sample
(i.e. numbers 2 and 19, 3 and 80, 5 and 10, 6 and 13, 11 and 14), there were
29 stars that were included twice in Tables 1 to 5.
These stars are listed, in order of
increasing right ascension, in Table 6. The table also includes the period,
the mean $V$ and $R$ magnitudes, denoted $<V>_F$ and $<R>_F$, because they
are the $A_0$ values derived from equation (1), and $\phi_{31}\pm \sigma$
for the $V$ data from each field
so that the two sets of observations
can be compared.
In the last column, we list the number of the table where all of the data
for the particular star can be found. It turns out that 24 of them 
are in Table 1.
Thus, although there are 680 entries in Table 1, they represent 656 different
stars. Our subsequent analysis is based on all 680 entries as the
duplicate entries do not sensibly alter the results.

Figure 1 is a plot of $<V>_F$ vs [$<V>_F-<R>_F$] for the data 
listed in Table 1.
For the bulk of the points, there is a general increase in $<V>$ with
increasing color, an expected  consequence of reddening. The
line shown in the diagram is the reddening vector  which has a
slope of $5.35$, the relative
extinction $A_V/E(V-R)$, for the Cerro Tololo
$V$ and $R$ bandpasses (Schlegel, Finkbeiner \& Davis  1998). 
The bright stars that appear in the upper right of the diagram are either
foreground or blended stars.
Their $V$ amplitudes, mean $V$ and $R$
magnitudes and colors are listed in Table 7. Since
blending of stars causes
the amplitude of light variation to be reduced, we assume that the stars with
$V$ amplitudes less than
$0.35$ are probably blended, but the ones with larger amplitudes
may be foreground stars. 
The 17 stars of Table 7 have been excluded from the rest of our investigation. 

In Table 8, we list the mean magnitudes and colors 
for the remaining program stars in each field. 
The angular coordinates of the field centers 
($\rho$ and $\Phi$)
are tabulated in columns (2) and (3).
Following van der Marel \& Cioni (2001, hereafter vMC01), we  define 
$\rho$ as the angular distance between the field center and the LMC center 
and $\Phi$ as the position angle of the field center measured eastward
from north. The origin of our adopted coordinate system 
is the location given by van der Marel (2001): 
$\alpha_0 =5^{\rm h} 29^{\rm m}$ and $\delta_0 =-69.^{\circ}5$.
According to vMC01,\footnote{vMC01
analysed near-infrared observations of stars at angular
distances ($\rho$) between 
$2.^{\circ}5$ and $6.^{\circ}7$ from the LMC center and found a sinusoidal
variation in brightness as a function of position angle. The peak to peak 
variation that they detected ($\sim 0.25$ mag) was attributed to
variations in the distance.} the inclination angle of the plane of the LMC 
disk is $34.^{\circ}7\pm 6.^{\circ}2$ and the line of
nodes has a position angle $\Theta=122.^{\circ}5\pm 8.^{\circ}3$, with the
near side at position angle $\Theta -90^{\circ}$ ($\Phi \sim 30^{\circ}$)
and the far side at $\Theta +90^{\circ}$ ($\Phi \sim 210^{\circ}$). 
Using these values along with equation (8) of vMC01, we calculated
for each field, the distance $D$ from the observer to the point where 
the field center
intersects the plane of the LMC disk, in terms of $D_0$, the distance
from the observer to the LMC center.  These $D/D_0$ values are
tabulated in column (4).
In column (5), we list the average densities of the fields 
(number of objects per square arcmin) that
were estimated by Alcock et al. (2001).
In columns (6) to (8), we list the mean $<V>_F$ and mean $<R>_F$ 
magnitudes and the mean colors
($<V>_F-<R>_F$) along with their standard deviations. The number of
program stars in each field is in column (9). 
The mean $V$ magnitudes range from $19.24$ for field \#19 to
$19.53$ for field \#15. 
These variations  may be due to a combination of
differences in distance, reddening or
the intrinsic properties of the stars among the different fields.
Any differences due to calibration are expected to be small. A99
found an internal precision of
$\sigma _V=0.021$, $\sigma _R=0.019$ and $\sigma _{V-R}=0.028$ for stars 
(with $V< 18$) in overlapping fields and it appears that this precision
extends to fainter magnitudes. Among the 29 RR Lyrae variables
listed in Table 6, 17 are included in both field \#6 and \#13. The mean
$\Delta <V>_F$ for these stars is $0.012$ and the mean $\Delta <R>_F$ is 
$0.022$.
Since the LMC is inclined to the plane of the
sky, some of our fields must be closer than others. 
According to the vMC01  model, we would expect the mean magnitudes
of the RR Lyrae stars in
the MACHO fields that are closest to us, fields \#3 and \#82, to be
approximately $0.06$ mag brighter than the ones in the 
most distant fields, \#10 and \#13. However, this is not indicated by 
the data of Table 8.
Differences in extinction seem to be more important.
The fields with the
faintest mean $<V>$ magnitudes (\#3 and \#15)  
have mean colors that are redder than most of the
other fields. If the 
distribution of unreddened colors of the RR Lyrae stars 
in these two fields is similar to the other fields, higher extinction 
can account for their faint mean magnitudes. We will discuss the effect of
extinction in section 4.
Another source of the variations may be inhomogeneities in the properties of
the stars themselves. This is a problem we will address using Fourier
decomposition.

In Figure 2, we plot the Fourier phase differences $\phi_{21}$,
$\phi_{31}$ and $\phi_{41}$ versus $\log P$ for the $V$ data. 
The points are plotted as open circles with three different sizes to 
denote different error levels: the larger the size, the smaller the
error.
In general, the phase differences increase with 
increasing period. 
Some of the outliers, the 
stars with $\phi_{31}\sim 3.5$, $\phi_{41}\sim 2.0$ and
$\log P < -0.55$ are probably second overtone pulsators (RR2 variables).
A96 have already pointed out
that there may be second overtone pulsators among
the LMC RR Lyrae population. 
The histograms of Figure 3 illustrate the range of errors
in the Fourier phase differences. As expected, the errors in $\phi_{41}$
are larger than those for $\phi_{31}$ which in turn are larger than the
errors in $\phi_{21}$. This occurs because the amplitudes for the higher 
orders are smaller and thus it
becomes increasingly difficult to derive their phases with sufficient
precision.
Figure 4 illustrates the relationship between the $\phi_{21}$, $\phi_{31}$ and
$\phi_{41}$ values for R and V. Horizontal lines are drawn at 
$\Delta \phi_{j1}=0$
on each plot. Although there is a great deal of scatter, it can
be noted that, in each case, the
majority of the points lie above the line. The mean differences
$<[\phi_{j1} (R) - \phi_{j1} (V)]>$ are $0.03$, $0.07$ and $0.10$ for j=2, 3
and 4 respectively,
indicating that in general, the Fourier phase differences for $R$ magnitudes
are greater than the ones for $V$. In a  comparison of $\phi_{21}$ and
$\phi_{31}$ for classical Cepheids,
Simon \& Moffett (1985) obtained a similar result. They found that 
$\phi_{j1}(R) > \phi_{j1}(V) > \phi_{j1}(B)$ for j=2 and 3. They also 
found that the differences were
greater for $\phi_{31}$ than for $\phi_{21}$.

In Figure 5, we plot the $V$ amplitude,  the Fourier amplitude $A_1$ and the 
amplitude ratios $R_{21}$,
$R_{31}$ and $R_{41}$ versus $\log P$ for the $V$ data. 
The  distributions of the
estimated errors for
$R_{21}$, $R_{31}$ and $R_{41}$ are shown in Figure 6. The error distribution
is similar for all three, but since the $R_{31}$ and $R_{41}$ ratios are 
significantly lower than $R_{21}$, their errors are relatively large.
Figure 5 illustrates that, in general,
$R_{21}$ decreases with increasing period, but there
is not such a clear trend for $R_{31}$ or $R_{41}$, possibly because
of their larger uncertainties. 
Some of the short period variables with low values for
$A_V$, $A_1$ and $R_{21}$ are probably RR2 variables.
Figure 7 shows the amplitude ratios ($R/V$) for the light curve
amplitude and $A_1$ through $A_4$. $R$ amplitudes are generally lower
than $V$ amplitudes because 
pulsating stars like RR Lyrae variables have lower
amplitudes when observed at longer wavelengths. 
In each panel, horizontal lines indicating the median
are shown, and in each case, the median $R/V$ ratio is approximately
$0.8$. The scatter is greater for the
higher orders because their amplitudes are small and have large uncertainties. 
Another factor that may contribute to the scatter in the amplitude
ratios is contamination. The presence of a nearby unresolved companion
will reduce the observed amplitude. This is an effect we need to
consider when we select a sample of stars for deriving the LMC distance.

\subsection{Comparision with the RR1 Variables in Galactic Globular Clusters}

For this comparison, we examine the plots of
$\phi_{31}- \log P$ and $R_{21} -\log P$. We prefer $\phi_{31}$ to
$\phi_{21}$, even though $\sigma (\phi_{31}) > \sigma (\phi_{21})$
because the range in values of $\phi_{21}$ is very small
making it difficult to detect the differences among the clusters.
The $\phi_{31}-\log P$ plots are shown in Figure 8.
In the upper panel,
we plot $\phi_{31}(V)$ vs $\log P$ for the RR1 variables in
five well-studied GGCs. The data for
these clusters are taken
from the following sources: NGC 6441 (Pritzl et al. 2001), M107 (Clement 
\& Shelton 1997), M5 (Ka{\l}u\.{z}ny
et al. 2000, hereafter K00), M2 (Lee \& 
Carney 1999, hereafter LC99) and
M68 (Walker 1994). 
For NGC 6441, we have included all the variables that Pritzl et al.
(2001) listed as
`RRc' but not the questionable ones (indicated as `RRc?'). V79 
was also excluded because it had a large amount of 
scatter on its light curve. For M5, we have included all of the variables in
Table 1 of the K00 paper with the exception of V78 (considered to be an RR2
variable), V76 (classification uncertain) and V130 (large error in
$\phi_{31}$). Since the M5 Fourier decomposition was based on a sine series,
we subtracted $3.14$ from all of the published $\phi_{31}$ values before 
plotting them in Figure 8.
The LC99 study of M2 did not include Fourier analysis of the variables
so we analysed their published observations. 
Our results for the three M2 stars that we 
consider to be bona fide\footnote{LC99 presented photometry
for 30 RR Lyrae variables, 12 of which they classified as type c.
We include only three of these `type c' stars in Table 9 because three of them
(LC 651, 715 and 733) appear to be pulsating in the second overtone
mode. Four others (V15, V18, V20 and LC 939) were excluded because their 
light curves have night-to-night variations similar to those
of the variables in M55 that Olech et al. (1999) classified
as non-radial pulsators. The stars LC 608 and 1047 were also excluded 
because of a large amount of scatter on their light curves. In the case
of the former, it may be caused by large period changes and for the
latter, it may be due to light contamination from a nearby star.} RR1 
variables are summarized in Table 9. For 
M68, we excluded V5 which is probably an RR2 variable.
The straight lines in the upper panel
are least squares fits to the data for each cluster.
In the central panel, these lines are plotted again, along with
the $\phi_{31}(V)$ values for the LMC stars with
$\sigma(\phi_{31})<0.4$. The lower panel is a repeat of the central panel,
but it also includes points 
plotted for the RR1 variables in M3 from 
Ka{\l}u\.{z}ny et al.
(1998)
and $\omega$ Centauri from the Clement \& Rowe (2000) study
based on the observations of 
Ka{\l}u\.{z}ny et al. (1997). 
The $\phi_{31}-\log P$ plots for the GGCs indicate that in general,
the higher the metallicity,\footnote {The
[Fe/H] values that Harris (1996) list in his 2003
catalog update for NGC 6441, M107, M5, M3, M2 and M68
are $-0.53$, $-1.04$, $-1.27$, $-1.57$, $-1.62$ and $-2.06$, respectively. The
ZW values for the same six clusters are $-0.59$, $-0.99$,
$-1.40$, $-1.66$, $-1.62$ and $-2.09$.}
the higher the line lies in the diagram. 
Most of the LMC RR1 variables are
distributed between the M107 and M68 lines. 
Many have $\phi_{31}-\log P$ values similar to the ones in M2,
M3, M5
and the Oosterhoff type I (OoI) variables 
in $\omega$ Centauri,  i.e. the ones with $\log P<-0.44$. However, not
many are similar to the OoII variables in $\omega$ Cen, the ones
with longer periods.
There are also a significant number (with
$\phi_{31}$ between $2.0$ and $3.0$) 
that do not have counterparts among the RR1 variables in the
well-studied GGCs. These objects are worthy of further investigation.

The $R_{21}-\log P$ plots are shown in Figure 9.
In the upper panel, we plot the data for the RR1
variables in M107, M5, M2 and M68. Lines based on least squares fits
to the points are plotted for M5, M2 and M68. 
However, we do not plot a line for
M107 because, although the five stars with the shortest periods show a
steady decrease in $R_{21}$ with increasing period, the other two (with
$\log P\sim -0.5$) do not follow the sequence. Instead, they lie among the
M5 points. In the central panel, these lines are plotted again, along with the
LMC variables. The lower panel is the same as 
the central panel, but includes points for M3 and
$\omega$ Centauri. Like the previous figure,
Figure 9 illustrates that some of the LMC RR1 variables
have $R_{21}-\log P$ values similar to the ones in M2, M3, M5 and 
the OoI variables in $\omega$ Cen, but they are different from the
majority of the $\omega$ Cen  variables which have OoII characteristics.

Since one of the aims of our investigation is to 
determine the LMC distance, we want to select 
a group of LMC
variables that have properties  similar to the variables in a 
well-studied GGC. We will then make 
the assumption that these stars have a similar
distribution of luminosities so that their absolute magnitudes can
be derived by independent methods. We conclude that
M5 is more suitable for this purpose than M2 or M3 
because the M5 study by K00 includes a more complete
sample (13)
of RR1 variables with well-determined Fourier coefficients. 
In addition, there have been independent studies of the absolute
magnitude of the HB stars in M5. Storm, Carney \& Latham  (1994) performed a
Baade-Wesselink analysis on two of its RR Lyrae variables
and Carretta et al. (2000) derived $M_V(HB)$ from main
sequence fitting. These  studies will be discussed further
in section $5.3$.

\subsection{Luminosity, Mass and Temperature Derived from Fourier Coefficients}

SC93 derived an equation, based on hydrodynamic pulsation models,
for calculating the 
luminosity of an RR1 variable from $\phi_{31}$ and period:
\begin{eqnarray}
\log L/\lsun=1.04\log P -0.058\phi_{31} +2.41  
\end{eqnarray}
In an independent investigation, based on observations of 93 RR1
variables in eight different stellar systems, Kov\'acs (1998, hereafter
K98) derived an equation relating the 
absolute magnitude $M_V$ 
of an RR1 variable to its period and Fourier coefficients $\phi_{21}$
and $A_4$: 
\begin{eqnarray}
M_V=1.261-0.961 P -0.044\phi_{21} -4.447A_4.
\end{eqnarray}
\noindent 
The phase difference $\phi_{21}$ in the K98 equation 
is  based on a sine series fit
so we subtracted $1.57$ from our  $\phi_{21}$ values which are based 
on a cosine series.
SC93 also used their pulsation models to derive an equation relating 
mass to $\phi_{31}$ and period: 
\begin{eqnarray}
\log M/\msun=0.52 \log P - 0.11 \phi_{31} +0.39.
\end{eqnarray}
Combining this with the period/mean density law (equation 2 in their
paper),
we can derive an 
equation for calculating the temperature:
\begin{eqnarray}
\log T_{eff}=3.775 -0.1452 \log P + 0.0056 \phi_{31}.
\end{eqnarray}
Using equations (2), (3) (4) and (5) with the $V$ data, we calculated 
$\log L$, $M_V$, $\log M$ and $\log T_{eff}$ for the `Table 1' stars with
$\sigma (\phi _{31})<0.4$. 
Since these equations are valid only for RR1 variables,
we have included only 
stars with periods in the range $-0.56 < \log P < -0.4$ and amplitudes
$A_V>0.3$. Stars with shorter periods and lower amplitudes are probably
RR2 variables and stars with longer periods have anomalous
light curves indicating that they are probably not bona fide RR1 variables. 
In addition, we restricted the sample 
to stars with amplitude ratios in the range:
$0.75 <A_R/A_V <0.85$. 
Amplitude ratios outside this range may occur
if the phase coverage on the light curve is incomplete or if the
star has a faint unresolved companion, in which case the mean magnitude
and amplitude are not reliable. 
Also, if a variable is an eclipsing binary, its amplitude ratio should
be close to unity and therefore $>0.85$. 
A total of 330 stars met the above criteria. We will refer to these as the
`bona fide' RR1 stars. In column (2) of the electronic version of Table 1, 
these stars are denoted `bf'.

Figure 10 is a plot of $\log L/\lsun$ versus $\log T_{eff}$ for these
330 bona fide RR1 variables. It demonstrates 
that the stars in our sample have $\log L$ ranging from $\sim 1.6$ to 
$\sim 1.8 \lsun$ and $\log T_{eff}$ between $\sim 3.85$ and $\sim 3.87$,
values appropriate for first-overtone pulsators on the blue side
of the instability strip according to the models of Bono et al. \rm
(1997) and Yoon \& Lee (2002).
In Figures 11 and 12, we plot $\log L$ and $M_V$ versus $<V>_F$ 
with the results
for each of the 16 fields shown in different panels. The lines in
Figure 11 have a slope of $-0.4$. They are plotted at an arbitrary position,
but are set at the same
position in each panel so that any variations
among the different fields can be readily recognized.
The standard deviation of the fit of the hydrodynamic models to equation (2) 
was $\Delta \log L=0.035$, but the scatter for the individual fields is 
greater than that. 
The situation is similar in Figure 12 where the lines are plotted with a slope
of unity. The standard deviation that K98 derived for equation
(3) was $0.042$.  
Clement \& Rowe (2000) made similar plots for the RR1 variables in
$\omega$ Centauri and the fit was much better. 
If the variations in apparent magnitude among the stars in the
individual LMC fields are due primarily to differences in luminosity,
we should see better correlations in Figures 11 and 12.
On the other hand, if the variations are due to differences in
distance as well as luminosity, 
we would not expect to see correlations in these figures.
However, a difference of $0.5$ mag would require a difference of
about 25\% in distance which is certainly not expected a priori.
We conclude that
differential reddening and crowding within the individual fields must
also contribute to the scatter. 

In Figure 13, we plot $\log L$ vs. $M_V$ and it is clear that these two 
quantities are correlated even though equations (2) and (3) were derived
by independent methods. 
Since C03 established that there exists a
luminosity-metallicity relation with slope =
$\Delta M_V/\Delta \rm{Fe/H}=0.214\pm 0.047$ 
among the LMC RR Lyrae variables,
we assume that the brighter stars are more metal-poor than the
faint ones. 
This means that their bolometric corrections will be different and
since $\log L$ refers to the bolometric luminosity, we should take
this into account when comparing $\log L$ with $M_V$.
Bessell \& Germany (1999, hereafter BG99) showed the relationship 
between bolometric correction (BC) and $(V-R)$ color for four different
values of [Fe/H] (Figure 8 in their paper). From a comparison of BC for 
[Fe/H]=$-1.0$ and $-2.0$ at $(V-R)_0=0.14$ mag, a typical intrinsic color for
an RR1 variable, we see that 
$\Delta BC/\Delta \rm {[Fe/H]}\sim 0.03$. Combining this
with C03's slope for the luminosity-metallicity relation ($0.21$), we derive a 
slope of $-0.46$ for the $\log L/\lsun$ vs. $M_V$ plot.  
The envelope lines in the diagram are plotted with this slope 
and are separated by $\Delta \log L=0.07$, twice the standard deviation
in the fit of equation (2) to the models.  
The actual slope of the plotted points $(-0.53)$ is steeper
than $-0.46$.
Nevertheless,
72\% of the points lie between the envelope lines. 
In section $5.3$, we will derive a mean absolute magnitude for the
M5-like variables based on Fourier coefficients.

%
%************************ SECTION 4
%

\section{THE INTERSTELLAR EXTINCTION}

In their discussion of the LMC distance, 
Benedict et al. (2002) pointed out that 
the average extinction-corrected magnitude of RR Lyrae 
variables in the LMC remains a significant uncertainty. 
Establishing the 
effect of interstellar extinction on  the observed magnitudes of LMC stars
is a difficult problem because the amount of extinction
is not constant.
Schwering \& Israel (1991) found that 
the foreground reddening ranges from $E(B-V)=0.07$ to $0.17$ mag 
over the LMC surface.  Among
the 16 fields in our study, their $E(B-V)$ values range from 
approximately $0.07$ to $0.14$ mag.
Thus it is not appropriate to make one reddening correction for
all of the stars in our sample. It is more accurate to consider
each star separately. Therefore, our approach is to calculate  the effective
temperatures from equation (5)
and then use a color-temperature calibration
to derive the unreddened colors.
Equations relating $\log T_{eff}$ to $(V-R)_0$ have been derived by
BG99 and by Kov\'acs \& Walker (1999, hereafter KW99).
BG99's equations, which are based on model atmospheres of Castelli (1999), 
apply only for $\log g=2.5$ and
four different values of [Fe/H]. However, the temperature-color 
relation of KW99 is more general. They 
derived a linear expression (equation 10 in their paper)
relating the temperature to color,
 $\log g$ and [M/H] based on models of Castelli, Gratton \& Kurucz 
(1997): 
\begin{eqnarray}
\log T_{eff}=3.8997 -0.4892 (V-R)_0 +0.0113 \log g + 0.0013 \rm{[M/H]}
\end{eqnarray}
and they also derived an expression (equation 12 in their paper)
for estimating the gravity from mass,
temperature and the fundamental mode pulsation period:
\begin{eqnarray}
\log g=2.9383 +0.2297 \log M/\msun - 0.1098 \log T_{eff}-1.2185 \log P_0.
\end{eqnarray}
We have used  equations (4), (5), (6) and (7)
to derive the unreddened color for each star. In each case, we
assumed [M/H]=$-1.5$ and the fundamental 
mode pulsation period was computed from the overtone period $(P_1)$
using a period ratio, $P_1/P_0=0.7445$. 
A plot of
$\log T_{eff}$  versus ($<V>-<R>$) for the bona fide
RR1 variables in each of the 16 fields is shown in Figure 14.
The unreddened line in each panel is derived from equation (6) assuming
$\log g=2.9$\footnote{In order to plot the unreddened line in Figure 14, we 
calculated a mean value of $\log g$ for the 330 stars in our sample
based on the mean
$\log T_{eff}$ $(3.863$), the mean $\log M$ $(-0.2085)$ and 
$\log P_0=-0.3617$, 
which corresponds to the mean $\log P_1$ $(-0.4898)$.} and [M/H]=$-1.5$.
The diagram indicates that the color excess varies from 
field to field and also
within the individual fields. Thus differential reddening may be responsible
for at least some of the scatter in Figures 11 and 12. 
For each of the stars, we calculated the color excess and
the corrected mean $V$ magnitude:
\begin{eqnarray}
E(V-R)=<V>_F-<R>_F-(V-R)_0 
\end{eqnarray}
\begin{eqnarray}
V_0(F)=<V>_F-5.35E(V-R) 
\end{eqnarray}
In Table 10, we summarize the mean $<V>_F$, the mean extinction
$E(V-R)$ and the mean corrected magnitude $V_0(F)$ 
for the bona fide RR1 stars in each field.  
N is the number of RR1 stars in the field and in each case, the errors 
represent the standard error of the mean (i.e. the standard deviation
divided by $\sqrt {N}$). In addition to these errors,
there are systematic errors in $E(V-R)$ and $V_0$ because of the
uncertainties in the derivation of $\log T_{eff}$ and $(V-R)_0$. These
will be discussed in section $5.2$.

The temperature-color relations that BG99 derived apply only
for $\log g=2.5$ which is lower than the $2.9$ that we have assumed.
According to equation (6), the lower $\log g$ $(2.5)$ would decrease $(V-R)_0$
by $\sim 0.01$ and thus increase the derived extinction. 
If we compare the $(V-R)_0$ colors derived from the BG99
relations with those predicted by our equation (6) for the same
$\log g$ (i.e. $2.5$),
we find that the BG99 colors would be about $0.01$ less
for $T_{eff} \sim 7300$, a typical temperature for RR1 variables.
Thus the extinction we 
derive for the same $\log g$ is lower than that 
predicted by BG99's equations.

On the other hand, the
mean extinction we derive for our bona fide RR1 stars is 
larger than the value that C03 adopted for the same region. 
C03 observed RR Lyrae variables in two fields that 
overlap with our fields 6 and 13 for which we derived mean
$E(V-R)=0.09$ and $0.08$ mag, respectively. This is 
equivalent to $E(B-V)\sim 0.15$ mag. 
They used two methods to determine the color
excess for the stars in their sample. First, they used 
Sturch's (1966) method. For this, they compared the
observed $(B-V)$ colors of 62 RR0 (RRab) variables at minimum light with 
unreddened colors that were calculated from Walker's (1990) equation
that relates color to period and metal
abundance. The $E(B-V)$ values that they derived for their 
fields A and B were $0.133$ and $0.115$ mag, respectively. In their second
method, they
derived the mean $(B-V)$ colors for five RR Lyrae at the blue and five at
the red edge of the instability strip and compared these
with the colors Corwin and Carney (2001) observed for the instability
strip boundaries in the globular cluster M3. To account for any color
differences due to different metallicity, they applied a metallicity-color
shift relation derived by Walker (1998).
Using this method, they derived $E(B-V)=0.116$ mag for their field A 
and $0.086$ mag for field B, values that are lower than the ones they
derived from the Sturch method. They adopted the latter values
because independent studies of globular clusters have indicated that
Sturch's method gives $E(B-V)$ values that are about $0.02$ mag larger
than those determined by other techniques. The mean extinction we derived
for fields 6 and 13 is comparable to what C03 found from the Sturch 
method, but considerably larger than the values they finally selected 
for their distance determination.

Nevertheless,
our extinction values are comparable to what Udalski et al. (1999) 
adopted for the OGLE investigations
of LMC RR Lyrae variables. Two of their fields
(SC19 and SC20) overlap MACHO field 15. Initially, Udalski (1998)
listed $E(V-I)=0.26$ mag for these two fields because this was the 
value used for the study of red clump stars  (Udalski et al. 1998).
It was based on a photometric (UBV) and spectroscopic 
study by Oestreicher \& Schmidt-Kaler (1996). However, the region
in question was not actually observed by these latter authors. It lies
about two degrees south of a region where they found high extinction.
When Udalski et al. (1999) re-estimated the reddening for these fields
from observations of red clump stars, they found that the extinction had
been overestimated in the earlier study. Their revision indicates
$E(V-I)\sim 0.23$ mag, which
corresponds to $E(V-R)\sim 0.11$ mag, the mean extinction that
we list for field 15 in Table 10. 

Another bench mark for  our extinction values is the 
recent study of LMC bump Cepheids by Keller \& Wood (2002),
based on MACHO data. Using pulsation theory, they derived the extinction
for 20 stars, two of which were in field \#19. 
The mean $E(V-R)$ that they derived for these two stars,
$0.03$ mag, is the same 
mean that we have derived for this field. 
We can not consider this to be a conclusive result however, because
their sample is so small.

\section{THE M5-LIKE VARIABLES}

\subsection{The Selection of the M5-like Variables}

Only the M5-like stars will be considered
for our derivation of the LMC distance and we select these stars
according to their location in the $\phi_{31}-\log P$ plots. 
Points that lie less than $2 \sigma$ ($\Delta \phi_{31}=0.214$)
from the M5 line in Figure 8 are considered to be
M5-like. ($\sigma$ is the standard deviation of
the fit of the K00 $\phi_{31}$ values to the line.)
In Figure 15, we show $\phi_{31}-\log P$ plots for the individual fields
with the lines for M5 and M68 superimposed.
The stars that are plotted as crosses are the ones we classify as
`M5-like'. In column (2) of the electronic version of Table 1, these
stars are denoted `M'.
The mean values of $<V>_F$, extinction $E(V-R)$ and $V_0(F)$ 
for the M5-like stars in each field are listed in the
last three columns of Table 10. In each case, the quoted error is the
standard error of the mean. These errors have been
derived in the same manner as those listed in columns 3 to 5 of the
table.

\subsection{The Mean Apparent Magnitude of the M5-like Variables}

In order to determine an LMC distance, we must derive the mean
apparent magnitudes of the M5-like stars in our sample, 
correct these magnitudes
for the effects of interstellar extinction and assess the errors. 
The mean $<V>_F$ and $<R>_F$ magnitudes listed in Table 10 are the
mean $A_0$ values calculated from equation (1). 
The first step is
to determine how well the $A_0$ values represent the data. To do this,
we selected a subset of the M5-like stars, the ten in field 80,
and applied Efron's bootstrap method (Diaconis \& Efron 1983).
The results indicated that the mean error in $A_0(V)$
was $0.0026$ mag with a standard deviation of $0.0010$. The equivalent
numbers for $A_0(R)$ were $0.0034$ and $0.0014$ respectively. 
Thus, the errors in $A_0$ are very small. 
The next
step is to compare $A_0$ with the `intensity' mean 
magnitude ($<V>_{int}$) because in
most studies of pulsating variable stars, the intensity means are used. 
To make this comparison, we
calculated $<V>_{int}$ and $<R>_{int}$ for the same 10 stars and the 
results are presented in Table 11.
It turns out that
the $A_0$ values are generally about $0.01$ mag fainter than the
intensity means. We will take this into account when we
adopt our final value for the mean corrected apparent magnitude.

In order to derive the interstellar extinction for each star,
we need to determine the color.
The mean colors listed in Tables 10 and 11 were computed from ($<V>_F-<R>_F$).
However, Fernie (1990) found that 
a straight average of the color over the pulsation cycle
(e.~g. $<V-R>_{mag}$),
is a better indicator of the temperature. We computed
$<V-R>_{mag}$ for the M5-like variables in field 80
and these are also listed in Table 11. 
The colors derived by the two methods differ by less than
$0.01$ mag. Thus we
conclude that using $(<V>_F-<R>_F)$ to derive the extinction
does not introduce any systematic errors.
The main sources of error in $E(V-R)$ are errors in the 
derived temperatures 
and in the color-temperature relation. Any errors in the observed 
colors are assumed to be minor and random, so
that over the sample of 80 M5-like stars, they cancel out. On the other hand,
the other two effects are not random because the stars in our sample
have similar properties. 
Based on the errors in the fits of SC93's equations relating
$\log L$ and $\log M$ of the pulsation models to the period
and $\phi_{31}$, 
($\sigma _{\log L} = 0.035$ and $\sigma _{\log M} = 0.025$),
we estimate that the error in $\log T_{eff}$
calculated from equation (5) is $0.01$. 
This translates to an error of $0.02$ mag in $(V-R)_0$.  
Since the dependence of $\log T_{eff}$ on $\phi_{31}$ is very weak, 
the errors in $\phi_{31}$ do not make a significant contribution to
the errors in our derived temperatures.
The discussion of the 
different $\log T_{eff}$-color relations in section 4
indicates there is an additional uncertainty of $\sim 0.01$ mag
in our $(V-R)_0$ colors. 
Adding in quadrature, we estimate that the systematic error in the
color excess $E(V-R)$ is $0.022$ mag. Since $A_V=5.35 E(V-R)$, this
propagates to an error of $0.12$ mag in $V_0$. 

The distribution of
$V_0(F)$ for the RR1 variables is shown in Figure 16. 
All of the 330 bona fide RR1 variables have
been plotted, with the M5-like stars represented as solid areas. 
This diagram and the data in Table 10 both illustrate that
the mean $V_0$
of the M5-like stars does not differ significantly from that of all of
the RR1 variables. 
In Figure 16, each field
is plotted separately in case there are systematic field-to-field
variations.
Within some of the
individual fields, the range in $V_0$ for the M5-like stars
is $0.5$ mag or more. Based on the K00 study of M5, one
would expect a scatter of at least $0.1$ mag in $V_0$ among the M5-like
variables. Some of  the additional scatter must be due to 
variations in distance from the observer, but some is probably caused by
crowding.  
Artificial star tests in the MACHO fields support this
conclusion. The results from these tests will be discussed further
in section $5.4$ where we estimate the LMC distance using M5-like stars,
and include a correction for crowding bias.
Figure 17, a plot of $<V_0>_F$ versus  the density 
for each field, also illustrates that crowding affects our derived
$V_0$ values. One would
not expect a perfect correlation in Figure 17 because it is
possible for stars in low density fields to have unresolved
companions. Nevertheless, the faintest mean
$V_0$ $(\sim 19.1)$ occurs in low density fields, 
and the mean $V_0$ is $\sim 0.3$ mag brighter  in the highest
density fields where the crowding is expected to be the most severe.
Although we have
tried to overcome the crowding problem by restricting our sample to stars
with an amplitude ratio $A_R/A_V$ between $0.75$ and $0.85$, it
appears that it has not been completely eliminated. 

Another method for checking our adopted
mean magnitudes is to compare with other investigations.
Seven of the bona fide RR1 stars from Table 1 were included in C03's
study. Their mean magnitudes will be published in a forthcoming
paper by Di Fabrizio et al. (2003).  
In Table 12, we list 
the mean $<V>$ magnitudes 
derived for these stars from the two studies; for both groups, the means
that we list are arithmetic means. 
The average of the C03 mean magnitudes is $0.07$ mag fainter than
ours. 
It is more difficult to compare our results with those of Udalski et al.
(1999) because most of the OGLE observations were in the $I$ band and the 
mean magnitudes that they published were corrected for extinction.
In addition, the mean that they quoted included RR Lyrae variables
in four different LMC fields.
In Udalski's 1998 study, a mean $V_0=18.86$ mag
was adopted for 110 RR0 variables in these fields,
two of which overlap with our field 15. This mean was derived from
$\sim 65$ I and 6 V magnitudes for each star. Later, when more observations
were available for these stars ($\sim 140$ in the I-band and 20
in the V-band) and the extinction was revised, Udalski et al.
(1999) changed the mean $V_0$ to $18.94 \pm 0.04$. 
The mean $<V_0>$ that we list in Table 10 for the 13 stars 
in field 15 is $19.00\pm  0.13$.

\subsection{The Absolute Magnitudes of the M5-like Variables}

In our derivation of the absolute magnitude, we consider four
independent methods: Baade-Wesselink (B-W) analysis, main sequence fitting,
trigonometric parallax of the star RR Lyrae and Fourier analysis.

As we noted at the end of section $3.2$,
Storm et al. \ (1994) performed a B-W analysis of two
RR Lyrae variables (V8 and V28) in M5. They derived  $M_V=0.65$ and $0.67$
mag, 
using a value of $1.30$ for $p$, the conversion
factor between observed and true pulsation velocity. 
Later, Clementini et al. (1995)  revised these 
values to $M_V=0.52\pm 0.26$ and $0.54 \pm 0.26$ mag, by
assuming [Fe/H]$=-1.17$
and $p=1.38$. However, if the assumed metal abundance is less, e.g. 
the ZW value of $-1.40$ 
instead of $-1.17$, these stars would 
be $0.02$ mag fainter.
The adopted $p$-factor ($1.38$) is the value that gives the brightest
possible value of the luminosity and Fernley (1994) considered it to be
a more appropriate value than $1.30$. 
More recently, Cacciari et al. (2000) re-examined the calibration 
of B-W results and suggested that $p$ might be a few 
percent smaller than $1.38$. Based on all of this, we adopt 
$M_V(BW)=0.55 \pm 0.26$ mag for the M5 RR Lyrae variables.

Carretta et al (2000) derived $M_V=0.54\pm 0.09$ mag for the
HB of M5 by fitting the main sequence to subdwarfs with parallaxes determined
from Hipparcos.  However, 
Gratton et al. (2002, 2003) have subsequently
concluded that this is too bright. 
Using the ESO VLT telescopes, they made new spectroscopic
observations of subdwarfs in the Galactic field
and in the three globular clusters NGC 6397, NGC 6752 and 47 Tuc
and have revised the  metal abundances. 
They also derived new reddening values. 
The new data indicate that the absolute $V$ magnitude for the horizontal
branch of M5 may be as bright as $0.58$ or as faint as
$0.65$, depending on the assumed metal abundance. We
adopt $M_V(HB)=0.61 \pm 0.12$ mag.

Benedict et al. (2002) used HST
astrometry to derive a trigonometric 
parallax for RR Lyrae and then calculated its
absolute magnitude: $M_V=0.61^{-0.11}_{+0.10}$. Because of the
location of RR Lyrae in a period-amplitude plot, we assume that its
absolute magnitude is comparable to that of an M5 RR Lyrae variable.
It is well
known that the light curve of the star, RR Lyrae ($P=$0\day 567), is 
modulated with a longer
secondary period ($\sim 40$ days), 
i.e. it exhibits the Blazkho effect (Smith 1995).
However, according to Szeidl (1988), the maximum
light amplitude of a Blazhko variable always fits the period-amplitude
relation for singly periodic variables.  According to
some recently published observations of RR Lyrae (Smith et al. 2003),
its maximum $V$ amplitude is $\sim 0.9$ mag, which 
places it on the period-amplitude
relation that K00 plotted for the RR0 variables in M5. 
If there is a period-luminosity-amplitude
relation for RR0 variables 
as demonstrated by Sandage (1981),
then we may assume that the
absolute magnitude of RR Lyrae is comparable to that of an M5-like variable.
We adopt
$M_V(\rm{trig}\ \pi)=0.61^{-0.11}_{+0.10}$ mag. 

Equations (2) and (3), which were derived
by SC93 and K98 respectively,
show relationships between
Fourier coefficients and the luminosity of RR1 variables.
K98 based the zero-point for his equation  on absolute magnitudes
derived from B-W analyses of field and cluster RR Lyrae variables. 
We have already taken into account the B-W analyses of two RR Lyrae 
variables in M5, but  we can not assume that the other stars for
which B-W analyses have been carried out, and on which K98 based
his zero point, have  properties similar to the RR Lyrae in M5.
Therefore, we will not derive an absolute magnitude from
equation (3), we will consider only equation (2).
The mean $\log L/\lsun$ for the 80 M5-like stars in our sample is
$1.688$. 
Assuming that the Sun's $M_{bol}=4.74$ (BG99), we  
derive a mean $<M_{bol}>=0.52$ for these M5-like stars.
From the $BC-(V-R)_0$ plot of BG99, we read $BC=0.02$ at 
[Fe/H]$\sim -1.5$ and 
$(V-R)_0=0.14$ (a typical color for an RR1 variable). 
Since the standard deviation of the fit of equation (2) to SC93's
models was $\sigma _{\log L}=0.035$, 
the mean $M_V$ derived from $\phi_{31}$ is $0.50\pm 0.09$ mag.

Using a weighted average of these four values,
our final adopted mean $M_V$ for 
M5-like RR Lyrae variables is $0.56\pm 0.06$ mag.

\subsection{The LMC Distance}

The average $V_0(F)$ for the 80 M5-like stars is $18.91\pm 0.02$ where
the quoted error is the standard error of the mean. If the
$<V_0>$ values for the individual fields are corrected for $D/D_0$,
this is revised to $18.89\pm 0.02$. We emphasize that making this
correction assumes that the
RR Lyraes lie in the main disk of the LMC. Fortunately, 
since our stars lie near the center of the LMC,
the difference between the two values is small.
Because of the systematic difference between $A_0(V)$ and $<V>_{int}$, 
we subtract $0.01$ from $<V_0(F)>$ and adopt a mean $V_0$
of $18.88\pm 0.02$ for the 80 M5-like stars in our 16 fields.
In addition, there is a systematic error of $0.12$ mag due to
the uncertainty in $E(V-R)$.
We must also account for crowding bias which is known to exist 
in the MACHO data because of the artificial star tests.

As the final step in calculating the LMC distance,
we estimate a correction for the apparent brightness of
the RR1 stars using 
artificial star tests that were part of
the MACHO microlensing detection efficiency calculation
(Alcock et al.~2001).  
This calculation is fraught with difficulty
because many of the steps
in the process of selecting bona fide RR1 variables, for example 
the removal of severely blended stars,
cannot be modeled accurately.
Therefore, our approach
is to interpret the artificial star tests in
two different ways, and thus
derive two magnitude corrections
for the RR1 stars.  
The difference between the two estimates obtained
is taken to indicate the level of systematic error
associated with the final adopted magnitude correction, and hence also
with our LMC distance result.

The MACHO artificial star tests were made in 10 
regions of the LMC that span the range of different crowding
conditions found in the MACHO database.  The parameter that
we adopt as a measure of the crowding is the number of
objects detected per square arcminute,
or what we call the `Density' in Table~8 (see also Tables 1 \& 2 of
Alcock et al.~2001).  In the most crowded MACHO fields,
this `Density' may have been underestimated because it is more difficult
to detect the faint objects.
The correction that we seek, $\Delta V$, is the difference between
the input and recovered magnitudes of the artificial stars.
For our first method, we calculated the average $\Delta V$ for
stars in 4 (1-mag wide) bins
centered at $V$ =17.5, 18.5, 19.5, and 20.5 mag.  
A high order polynomial function was fit to these
binned data points in order to predict $\Delta V$
in terms of recovered magnitude and field density for
any value of these two parameters.  
Since the frequency-weighted mean density for the M5-like RR1 stars in
our sample is 200 objects/arcmin$^2$, we evaluated the function
for a density of 200 objects/arcmin$^2$ and derived
$\Delta V$ (recovered $-$ input) $ = -0.21$ mag at $V = 19.3$.
However, because we used an (unclipped) straight average for $\Delta V$
in each bin, heavily blended artificial stars may tend to inflate
the correction.
In fact this correction 
is probably an upper limit for the
RR1 stars in our sample because we have already
removed the most severely blended stars,
but these are given full weight when interpreting the artificial stars.
For our second estimate, we constructed one bin of artificial stars 
with $V$ = 19.1 to 19.5 mag and calculated the
median $\Delta V$ value at each of the 10 different field densities.
A second-order polynomial that was forced to pass
through zero in an empty field was fit to these
data.\footnote{For $V$ = 19.1 to 19.5 mag,
$\Delta V =  -6.06\times10^{-5} \cdot O - 2.34\times10^{-6} \cdot O^2$,
where $O$ is objects/arcmin$^2$.} In this
case, $\Delta V = -0.11 \pm 0.10$ mag at a density of 200 object/arcmin$^2$.
Since the median value is not significantly affected by 
outlying points like those due to severe blends, we believe that
this approach is more appropriate for our data.
When this correction is applied, our $V_0$ is revised to 
$18.99$ mag.  In section $5.2$ and Table 12, we showed that our mean
$V$ magnitudes for seven RR1 stars in fields \#6 and \#13 appear to 
be $\sim 0.07$ brighter than those of C03.
When the crowding correction for these two fields is applied to
the MACHO magnitudes for these stars, the mean $V$ magnitude for
the MACHO data is $0.03$ mag fainter than the C03 values, well within the
estimated errors.

There are two sources of systematic error in our estimate of $V_0$. The first
is the error in our derived $(V-R)_0$ color ($0.022$ mag) which
propagates to
an error of $0.12$ mag in $V_0$. The second is the error
in our crowding correction ($0.10$ mag). Combining these in quadrature,
we  estimate that the systematic error in $V_0$ is $0.16$ mag. Another
factor that should be considered is the effect of crowding on the derived
Fourier coefficients. To address this problem, 
SC93 performed simulations in which they added constant light to the observed
magnitudes at all phases on the light curves. They found that, although $A_0$
brightened, $\phi_{31}$ remained largely unaltered.  We
therefore conclude that crowding has not seriously affected our
selection of the `M5-like' stars.

The final mean LMC distance modulus that we obtain for the 80 M5-like
RR1 variables, based on $M_V=0.56$ and $V_0=18.99$,
is $\mu_{LMC} = 18.43 \pm 0.06$ (statistical)
$\pm 0.16$ (systematic).
Our analysis has illustrated that the 
two major impediments to the derivation of a precise LMC distance 
from RR Lyrae variables are the uncertainty in the extinction and 
the uncertainty in establishing the effects of light contamination
due to crowding. Near-infrared photometry (see, for example, the recent study
of the Reticulum cluster by Dall'Ora et al. 2003)
is an effective way to deal with the first problem. 
To address
the second problem, investigations like the SuperMACHO project 
which is based on observations obtained
with the 4-meter telescope at Cerro Tololo will provide
images with better resolution. Preliminary results indicate that the
median seeing of the CTIO images is a factor of three smaller than
the FWHM of the images that were obtained with the Mt. Stromlo 50-inch
telescope for the MACHO project.

\section{SUMMARY}

We have determined Fourier coefficients for 785 stars deemed to be
first-overtone RR Lyrae (RR1) variables according to their
periods, magnitudes and colors. We established that 330
of these stars are bona fide RR1 stars. By using a $\phi_{31}-\log P$
plot, we compared the LMC stars with the RR1 variables in some
well-studied  GGCs and found that they have properties
similar to the ones in M2, M3, M5 and the OoI variables in
$\omega$~Centauri.   However, they are different from the OoII variables
in $\omega$~Cen. In addition, there are a significant number that do not have
counterparts in the well-studied GGCs.

There are several problems that must be addressed in  deriving the LMC
distance from RR Lyrae variables. Perhaps the most important is the 
correction for differential extinction. We have shown that there is a
large range in the color excess from field to field ($0.03 < E(V-R) < 0.13$),
in good agreement with the results of other studies. We also found
a significant variation among the stars within individual fields.
These variations indicate that it is advantageous to correct for the
extinction on a star-by-star basis. We did this by computing the
temperature of each star from $\phi_{31}$ and $\log P$ and then using
the temperature to derive the unreddened color.

Another consideration is that the star fields in the LMC, particularly those
in the region of the bar, are extremely crowded. By comparing the
$V$ and $R$ pulsation amplitudes, we removed the most severely blended
stars from our sample. Further, a statistical correction based on artificial star
tests was made to rectify the photometric errors caused by more moderate
blending.

Finally, it is important to 
select a homogeneous group of stars for which the absolute magnitude
is well-determined.  To do this, we identified the `M5-like' RR1 stars
in our sample and applied absolute magnitudes determined from four
independent methods to derive a distance modulus of $18.43 \pm 0.06$
(statistical) $\pm 0.16$ (systematic) mag. This method has the
advantage that the result does not depend on the still somewhat uncertain
$M_V-\rm {[Fe/H]}$ relation for RR Lyrae variables.

%
%********************** ACKNOWLEDGEMENT
%
\acknowledgements

We are very grateful for the skilled support given our project by the 
technical staff at the Mt. Stromlo Observatory. 
In addition, it is a pleasure to thank Ed Guinan and Gisella Clementini 
for valuable discussions with CMC during the preparation of this paper.
Gisella responded cheerfully and promptly to the many questions
we asked and provided us with some of her unpublished results.
This work was performed
under the auspices of the U.S. Department of Energy, National Nuclear
Security Administration by the University of California, Lawrence
Livermore National Laboratory under contract No. W7405-ENG-48, the
National Science Foundation through the Centre for Particle Astrophysics
of the University of California under cooperative agreement AST-8809616,
and the Mount Stromlo and Siding Springs Observatory by the Bilateral
Science and Technology Program of the Australian Department of Industry,
Technology and Regional Development. 
CMC and DLW were supported by the Natural Sciences and Engineering 
Research Council of Canada (NSERC). AM and JFR held NSERC Undergraduate
Student Research Awards during this work.
KG and TV were supported in part by the Department of Energy under grant
DEF03-90-ER 40546.
DM is supported by FONDAP Center
for Astrophysics 15010003.

\newpage

\clearpage
\centerline{\bf{TABLE CAPTIONS}}

{\sc Table} 1. {\sc Fourier Parameters of LMC RR1 Variables}

{\sc Table} 2. {\sc Fourier Parameters of LMC RR01 Variables}

{\sc Table} 3. {\sc Fourier Parameters of LMC RR12 Variables}

{\sc Table} 4. {\sc Fourier Parameters of LMC RR1-$\nu 1$ Variables}

{\sc Table} 5. {\sc Fourier Parameters of Other Multifrequency LMC RR1 Variables}

{\sc Table} 6. {\sc LMC RR1 Variables Included in Two Fields}

{\sc Table} 7. {\sc Blended or Foreground RR1 Variables}

{\sc Table} 8. {\sc Mean Magnitudes and Colors of LMC RR1 Variables by
Field}

{\sc Table} 9. {\sc Fourier Parameters for the RR1 Variables in M2}

{\sc Table} 10. {\sc Corrected Magnitudes and Reddenings by Field}

{\sc Table} 11. {\sc Mean Magnitudes and Colors for the M5-like Stars in 
Field 80}

{\sc Table} 12. {\sc Mean Magnitudes for Stars Observed by C03}

%
%***********************TABLES

\clearpage

\begin{deluxetable}{cccccccccc}
%\footnotesize
\tablecolumns{10}
\tablecaption{Fourier Parameters of LMC RR1 Variables}
%\tablewidth{Opt}
\tablehead{\colhead{Color \& Star\#} & 
 \colhead{Period} & \colhead{$A_{0}$} & \colhead{$A_{1}$} & \colhead{$R_{21}$} 
& \colhead{$R_{31}$} & \colhead{$R_{41}$} &\colhead{$\phi_{21}$} 
&\colhead{$\phi_{31}$} &\colhead{$\phi_{41}$} \\
\colhead{$N_{obs}$} 
& \colhead{Amplitude} & \colhead{$\sigma_{fit}$} & \colhead{} &
\multicolumn{6}{l}{----------------------------- $\sigma$ -----------------------------} 
 \\
\colhead{(1)}& \colhead{(2)}& \colhead{(3)}& \colhead{(4)}& \colhead{(5)}&
\colhead{(6)}& \colhead{(7)}& \colhead{(8)}& \colhead{(9)}& \colhead{(10)}
}
\startdata

V80.6589.1879  &0.249987&19.24&0.149&0.1129&0.0324&0.0318&4.81&3.36&1.87\nl
    751   &0.304 & 0.066 & &0.0236& 0.0238& 0.0244& 0.22& 0.74& 0.75 \nl
  
R80.6589.1879 &0.249987&18.94&0.104&0.1198&0.0359&0.0546&4.56&2.39&2.16\nl
    736   &0.213 & 0.071 & &0.0368& 0.0371& 0.0378& 0.32& 1.05& 0.69 \nl
  
V2.5389.1478  &0.250557&19.52&0.178&0.0679&0.0611&0.0117&4.99&3.04&4.03\nl
    469    &0.358 &0.082 & &0.0303& 0.0310& 0.0299& 0.46& 0.51& 2.69 \nl
  
R2.5389.1478  &0.250557& 19.30& 0.136& 0.1228& 0.0481& 0.0561&4.57&3.54&4.59\nl
    455    &0.286 &0.074 & &0.0353& 0.0366& 0.0354& 0.32& 0.75& 0.67 \nl
  
V6.5849.1114  &0.255347&19.55&0.189&0.1462&0.0593&0.0161&4.39&3.43&3.23 \nl
    363    &0.391 &0.060 & &0.0247& 0.0235& 0.0239& 0.17& 0.41& 1.48 \nl
  
R6.5849.1114 &0.255347&19.36&0.147&0.1217&0.1044&0.0546&4.62&3.99&1.38\nl
    372      &0.284 &0.059 & &0.0302& 0.0298& 0.0293& 0.25& 0.30& 0.57\nl
  
\enddata
\tablecomments{Table 1 is available in its entirety in the 
electronic edition of the Astronomical Journal. A portion is shown here for 
guidance regarding its form and content.}
\end{deluxetable}

\begin{deluxetable}{cccccccccc}
%\footnotesize
\tablecolumns{10}
\tablecaption{Fourier Parameters of LMC RR01 Variables}
%\tablewidth{Opt}
\tablehead{\colhead{Color \& Star\#} & 
 \colhead{Period} & \colhead{$A_{0}$} & \colhead{$A_{1}$} & \colhead{$R_{21}$} 
& \colhead{$R_{31}$} & \colhead{$R_{41}$} &\colhead{$\phi_{21}$} 
&\colhead{$\phi_{31}$} &\colhead{$\phi_{41}$} \\
\colhead{$N_{obs}$} 
& \colhead{Amplitude} & \colhead{$\sigma_{fit}$} & \colhead{} &
\multicolumn{6}{l}{----------------------------- $\sigma$ -----------------------------} 
 \\
\colhead{(1)}& \colhead{(2)}& \colhead{(3)}& \colhead{(4)}& \colhead{(5)}&
\colhead{(6)}& \colhead{(7)}& \colhead{(8)}& \colhead{(9)}& \colhead{(10)}
}
\startdata

V80.7193.1485 &0.328817&19.50&0.215&0.1431&0.0766&0.0157&5.14&4.98&1.23\nl
    514  & 0.433& 0.084 & &0.0244& 0.0248& 0.0249& 0.18& 0.34& 1.62 \nl
  
R80.7193.1485 &0.328817&19.24&0.168&0.1701&0.1139&0.0528&4.99&5.33&5.55\nl
    394  & 0.379& 0.099 & &0.0409& 0.0430& 0.0428& 0.28& 0.40& 0.85 \nl
  
V81.8639.1450 &0.335260&19.14&0.124&0.1152&0.1500&0.0356&4.35&2.77&1.21\nl
    435  & 0.256& 0.069 & &0.0380& 0.0389& 0.0386& 0.34& 0.28& 1.07 \nl
  
R81.8639.1450 &0.335260&18.92&0.101&0.1169&0.0766&0.0296&4.21&3.10&4.66\nl
    466  & 0.203& 0.068 & &0.0449& 0.0455& 0.0447& 0.39& 0.59& 1.56 \nl
  
\enddata
\tablecomments{Table 2 is available in its entirety in the 
electronic edition of the Astronomical Journal. A portion is shown here for 
guidance regarding its form and content.}
\end{deluxetable}

\begin{deluxetable}{cccccccccc}
%\footnotesize
\tablecolumns{10}
\tablecaption{Fourier Parameters of LMC RR12 Variables}
%\tablewidth{Opt}
\tablehead{\colhead{Color \& Star\#} & 
 \colhead{Period} & \colhead{$A_{0}$} & \colhead{$A_{1}$} & \colhead{$R_{21}$} 
& \colhead{$R_{31}$} & \colhead{$R_{41}$} &\colhead{$\phi_{21}$} 
&\colhead{$\phi_{31}$} &\colhead{$\phi_{41}$} \\
\colhead{$N_{obs}$} 
& \colhead{Amplitude} & \colhead{$\sigma_{fit}$} & \colhead{} &
\multicolumn{6}{l}{----------------------------- $\sigma$ -----------------------------} 
 \\
\colhead{(1)}& \colhead{(2)}& \colhead{(3)}& \colhead{(4)}& \colhead{(5)}&
\colhead{(6)}& \colhead{(7)}& \colhead{(8)}& \colhead{(9)}& \colhead{(10)}
}
\startdata

V12.10443.367& 0.336557&18.91&0.099&0.1408&0.1096&0.0261&1.30&6.16&3.10\nl
    674   & 0.215& 0.056 & &0.0320& 0.0319& 0.0311& 0.23& 0.30& 1.22\nl
  
R12.10443.367& 0.336557&18.60&0.076&0.1588&0.1035&0.0541&1.73&0.29&3.80\nl
    725   & 0.174& 0.057 & &0.0400& 0.0406& 0.0399& 0.26& 0.39& 0.75\nl
  
V12.10202.285& 0.398113&18.76&0.111&0.2345&0.2008&0.0918&2.11&0.09&4.14\nl
    565   & 0.262& 0.056 & &0.0313& 0.0314& 0.0316& 0.15& 0.17& 0.34\nl
  
R12.10202.285& 0.398113&18.43&0.092&0.2535&0.2196&0.0843&2.19&0.32&4.42\nl
    601   & 0.216& 0.052 & &0.0341& 0.0343& 0.0344& 0.15& 0.17& 0.41\nl

\enddata
\end{deluxetable}

\begin{deluxetable}{cccccccccc}
%\footnotesize
\tablecolumns{10}
\tablecaption{Fourier Parameters of LMC RR1-$\nu$1 Variables}
%\tablewidth{Opt}
\tablehead{\colhead{Color \& Star\#} & 
 \colhead{Period} & \colhead{$A_{0}$} & \colhead{$A_{1}$} & \colhead{$R_{21}$} 
& \colhead{$R_{31}$} & \colhead{$R_{41}$} &\colhead{$\phi_{21}$} 
&\colhead{$\phi_{31}$} &\colhead{$\phi_{41}$} \\
\colhead{$N_{obs}$} & \colhead{Amplitude} 
& \colhead{$\sigma_{fit}$} & \colhead{} &
\multicolumn{6}{l}{----------------------------- $\sigma$ -----------------------------} 
 \\
\colhead{(1)}& \colhead{(2)}& \colhead{(3)}& \colhead{(4)}& \colhead{(5)}&
\colhead{(6)}& \colhead{(7)}& \colhead{(8)}& \colhead{(9)}& \colhead{(10)}
}
\startdata

V14.9702.401  &0.275403&19.39&0.224&0.1919&0.0823&0.0562&4.50&2.62&1.42\nl
    532     &0.455 &0.067 & &0.0187& 0.0181& 0.0182& 0.10& 0.24& 0.34\nl
  
R14.9702.401  &0.275403&19.23&0.183&0.2491&0.0787&0.0460&4.51&2.52&1.18\nl
    539     &0.389 &0.063 & &0.0216& 0.0207& 0.0209& 0.09& 0.28& 0.47\nl
  
V6.5730.4057  &0.276320&19.41&0.117&0.2582&0.0262&0.0478&4.18&2.61&2.10\nl
    519     &0.262 &0.075 & &0.0416& 0.0415& 0.0402& 0.19& 1.53& 0.88\nl
  
R6.5730.4057  &0.276320&19.30&0.100&0.2406&0.0803&0.0455&4.25&2.14&1.76\nl
    483     &0.218 &0.068 & &0.0450& 0.0451& 0.0446& 0.22& 0.57& 1.01\nl
  
\enddata
\tablecomments{Table 4 is available in its entirety in the 
electronic edition of the Astronomical Journal. A portion is shown here for 
guidance regarding its form and content.}
\end{deluxetable}

\begin{deluxetable}{cccccccccc}
%\footnotesize
\tablecolumns{10}
\tablecaption{Fourier Parameters of Other Multifrequency LMC RR1 Variables }
%\tablewidth{Opt}
\tablehead{\colhead{Color \& Star\#} & 
 \colhead{Period} & \colhead{$A_{0}$} & \colhead{$A_{1}$} & \colhead{$R_{21}$} 
& \colhead{$R_{31}$} & \colhead{$R_{41}$} &\colhead{$\phi_{21}$} 
&\colhead{$\phi_{31}$} &\colhead{$\phi_{41}$} \\
\colhead{$N_{obs}$} 
& \colhead{Amplitude} & \colhead{$\sigma_{fit}$} & \colhead{} &
\multicolumn{6}{l}{----------------------------- $\sigma$ -----------------------------} 
 \\
\colhead{(1)}& \colhead{(2)}& \colhead{(3)}& \colhead{(4)}& \colhead{(5)}&
\colhead{(6)}& \colhead{(7)}& \colhead{(8)}& \colhead{(9)}& \colhead{(10)}
}
\startdata

V3.6243.404  &0.270850&19.41&0.196&0.2331&0.0761&0.2712&4.64&3.29&2.23\nl
    128     &0.467 &0.101 & &0.0826&0.0812& 0.0738& 0.33& 1.07& 0.47\nl
  
R3.6243.404  &0.270850&19.19&0.168&0.1906&0.1450&0.1716&4.93&2.38&1.88\nl
    189     &0.362 &0.085 & &0.0548&0.0563&0.0562&0.29&0.42&0.37\nl

V80.7441.933 &0.273310&19.28&0.165&0.1864&0.0331&0.0491&4.61&3.37&2.49\nl
    369     &0.349 &0.098 & &0.0464& 0.0446& 0.0441& 0.25& 1.33& 0.92\nl
  
R80.7441.933 &0.273310&19.04&0.141&0.1653&0.1261&0.0518&4.62&3.34&2.80\nl
    359     &0.292 &0.068 & &0.0376& 0.0363& 0.0359& 0.23& 0.31& 0.72\nl
  
\enddata
\tablecomments{Table 5 is available in its entirety in the 
electronic edition of the Astronomical Journal. A portion is shown here for 
guidance regarding its form and content.}
\end{deluxetable}

\begin{deluxetable}{lccccccc}
%\footnotesize
\tablecaption{RR Lyrae Stars Included in Two Fields }
\tablewidth{0pt}
\tablehead{
\colhead{Star} & \colhead{RA} & \colhead{DEC} 
& \colhead{Period} & $<V>_F$  
& $<R>_F$ & $\phi_{31}\pm\sigma$ 
& \colhead{Table no.} \\
\colhead{} & \colhead{}& \colhead{} & \colhead{(days)} 
& \colhead{} & \colhead{}& \colhead{(V)}   
& \colhead{}   
}
\startdata
    
10.4400.4594 & 05:07:39.804 & -69:58:26.57 & 0.332233 & 
19.26 & 19.04 & $3.63\pm0.33$ & 1 \nl
5.4400.1081 & 05:07:40.150 & -69:58:27.11 & 0.332233 & 
19.24 & 19.06 & $4.19\pm0.30$ & 1 \nl
 \nl
                                              
19.4789.4120& 05:09:39.150&  -68:15:00.80 & 0.326913 &
19.32&  19.13&  $3.60\pm 0.35$ &1 \nl
 2.4789.946 & 05:09:39.189&  -68:15:00.44 & 0.326910 &
 19.23&  19.05&  $3.20\pm  0.17$ &1 \nl
\nl

 6.5721.289 &  05:15:37.913&  -70:37:50.48&  0.335580 &
 19.14&  18.93& $2.90 \pm  0.19$&1 \nl
13.5721.2108&  05:15:38.028&  -70:37:50.15&  0.335580 &
19.16&  18.94&  $2.59\pm  0.20$ & 1 \nl
\nl

13.5965.3725&  05:17:08.129&  -70:31:19.35&  0.316960 &
19.31&  19.07&  $2.70\pm  0.27$&1 \nl
 6.5965.835 &  05:17:08.279&  -70:31:19.29&  0.316963 &
 19.32&  19.11& $ 3.12\pm  0.21$&1 \nl
\nl

13.6084.2519&  05:18:09.290&  -70:38:12.73&  0.306750 &
19.40&  19.15&  $2.24\pm  0.24$&1 \nl
 6.6084.462 &  05:18:09.305&  -70:38:12.43&  0.306747 &
 19.41&  19.15& $ 2.78\pm  0.23$&1  \nl
\nl

13.6569.4284&  05:21:14.235&  -70:34:00.52&  0.315017 &
19.45&  19.23&  $3.29\pm  0.28$&1  \nl
 6.6569.955 &  05:21:14.199&  -70:34:00.34&  0.315017 &
 19.44&  19.23& $ 2.83\pm  0.19$&1  \nl
\nl
                                                        
13.6689.3055&  05:21:33.252&  -70:39:52.25&  0.305057 &
19.25&  19.03&  $2.54\pm 0.22$&1  \nl
 6.6689.563 &  05:21:33.487&  -70:39:51.81&  0.305057 &
 19.24&  19.03& $ 2.40\pm 0.15$&1 \nl
\nl

80.6839.4533&  05:22:22.451&  -68:44:58.54&  0.304307 &
19.77&  19.52&  $2.58\pm 0.11$&1   \nl
 3.6839.2292&  05:22:22.497&  -68:44:57.87&  0.304307 &
 19.75&  19.52& $ 2.63\pm 0.20$&1   \nl
\nl

80.7201.3035&  05:24:42.268&  -68:47:00.97&   0.319110 &
19.12&  18.84&  $2.18\pm 0.16$&1  \nl
 3.7201.504 &  05:24:42.314&  -68:47:01.52&   0.319110 &
 19.24&  18.95& $ 2.24\pm 0.19$&1   \nl
\nl

11.8744.752 &  05:33:59.600&  -70:48:10.99&  0.280960 &
19.19&  19.05&  $2.60\pm 0.35$&1  \nl
14.8744.3856&  05:33:59.777&  -70:48:13.29&  0.280957 &
19.22&  19.07&  $2.68\pm 0.27$&1   \nl
\nl
                                                
11.8863.163 &  05:35:05.584&  -70:54:04.95&  0.331207 &
19.56&  19.40&  $2.51\pm 0.13$&1  \nl
14.8863.1362&  05:35:05.812&  -70:54:05.05&  0.331203 &
19.54&  19.37&  $2.53\pm 0.23$&1   \nl
\nl

\enddata
\tablecomments{$<V>_F$ and $<R>_F$ denote $A_0(V)$ and $A_0(R)$
which were derived from the fits of equation (1) to the observations.
Table 6 is available in its entirety in the electronic edition of the
Astronomical Journal. A portion is shown here for guidance regarding 
its form and content. The 11 pairs of stars listed above are all
considered to be bona fide RR1 variables.}
\end{deluxetable}

\begin{deluxetable}{lccccc}
%\footnotesize
\tablecolumns{6}
\tablecaption{Blended or Foreground RR1 Variables}
%\tablewidth{Opt}
\tablehead{\colhead{Star} & 
 \colhead{Period} & \colhead{$A(V)$} & \colhead{$<V>_F$} & \colhead{$<R>_F$} 
 & \colhead{$<V>_F-<R>_F$} 
 \\
\colhead{(1)}& \colhead{(2)}& \colhead{(3)}& \colhead{(4)}& \colhead{(5)}& 
\colhead{(6)}
}
\startdata

 2.5873.332  & 0.2965 & 0.27 & 18.74 & 18.40 & 0.34 \nl
 2.5148.713  & 0.3213 & 0.34 & 18.96 & 18.65 & 0.31 \nl
 2.5752.387  & 0.4592 & 0.37 & 18.66 & 18.37 & 0.29 \nl
 3.7450.214  & 0.3919 & 0.26 & 18.98 & 18.65 & 0.33 \nl
 5.5252.708  & 0.3663 & 0.39 & 18.81 & 18.53 & 0.28 \nl
 6.6576.558  & 0.2692 & 0.20 & 18.74 & 18.37 & 0.37 \nl
 6.5851.3773 & 0.3210 & 0.35 & 18.77 & 18.45 & 0.32 \nl
 6.6811.481  & 0.3560 & 0.35 & 18.86 & 18.56 & 0.30 \nl
 11.9111.591 & 0.3146 & 0.24 & 19.04 & 18.69 & 0.35 \nl
 13.6438.48  & 0.2937 & 0.27 & 18.96 & 18.60 & 0.36 \nl 
 19.4303.852 & 0.2617 & 0.24 & 18.76 & 18.46 & 0.30 \nl
 80.6950.6196& 0.3022 & 0.43 & 19.07 & 18.73 & 0.34 \nl
 80.7315.1237& 0.3191 & 0.30 & 18.92 & 18.62 & 0.30 \nl
 80.7313.3932& 0.3406 & 0.42 & 19.10 & 18.73 & 0.37 \nl
 80.7192.4319& 0.3496 & 0.32 & 19.02 & 18.61 & 0.41 \nl
 80.7436.1309& 0.3692 & 0.29 & 18.89 & 18.59 & 0.30 \nl
 81.8639.662 & 0.2853 & 0.26 & 18.70 & 18.35 & 0.35 \nl

\enddata
\end{deluxetable}

\begin{deluxetable}{ccrcccccc}
%\footnotesize
\tablecolumns{9}
\tablecaption{Mean Magnitudes and Colors of LMC RR1 Variables by Field}
%\tablewidth{Opt}
\tablehead{\colhead{Field} & \colhead{$\rho$} & \colhead{$\Phi$} &
\colhead{$D/D_0$} & \colhead{Density} &
\colhead{$<V>_F$} & \colhead{$<R>_F$} & \colhead{Color}& 
  \colhead{$N_{RR}$}  \\
 &&&&&\multicolumn{3}{l}{------------------ Mean/$\sigma$ ------------------}& 
 \\
\colhead{(1)}& \colhead{(2)}& \colhead{(3)}& \colhead{(4)}&\colhead{(5)}  
&\colhead{(6)} &\colhead{(7)} &\colhead{(8)} &\colhead{(9)}
}
\startdata
2  &1.77$^\circ$& 304$^\circ$& 0.9999 & 184.7& 
  $19.30/0.21$ &$19.09/0.20$& $0.21/0.04$&66\nl

3  &1.21$^\circ$& 331$^\circ$& 0.9933 & 173.4& 
  $19.50/0.25$ &$19.22/0.24$& $0.27/0.04$&37\nl

5  &1.56$^\circ$& 262$^\circ$& 1.0128 & 237.9& 
  $19.26/0.18$ &$19.06/0.17$& $0.20/0.04$&55\nl

6  &1.12$^\circ$& 223$^\circ$& 1.0137 & 226.4& 
  $19.36/0.18$ &$19.14/0.18$& $0.22/0.05$&84\nl

10 &2.17$^\circ$& 257$^\circ$ & 1.0198 & 182.3& 
  $19.30/0.18$ &$19.09/0.17$& $0.21/0.03$&34\nl 

11 &1.25$^\circ$& 147$^\circ$ & 1.0065 & 221.9& 
  $19.41/0.23$ &$19.22/0.21$& $0.20/0.06$&43\nl

12 & 1.79$^\circ$& 129$^\circ$ & 1.0029 & 202.9& 
  $19.40/0.17$ &$19.16/0.16$& $0.24/0.04$&22\nl

13 &1.58$^\circ$& 209$^\circ$ & 1.0198 &  179.6&
  $19.33/0.14$ &$19.11/0.13$& $0.22/0.04$&78\nl

14 &1.76$^\circ$& 161$^\circ$ & 1.0139 & 211.1& 
  $19.29/0.15$ &$19.11/0.14$& $0.18/0.04$&27\nl

15 &2.25$^\circ$& 143$^\circ$ & 1.0104 & 182.9& 
  $19.53/0.14$ &$19.28/0.14$& $0.25/0.04$&24\nl

18 &2.82$^\circ$& 278$^\circ$ & 1.0156 & 214.0&
  $19.29/0.18$ &$19.09/0.18$& $0.20/0.04$&22\nl

19 &2.37$^\circ$& 297$^\circ$ & 1.0036 & 169.6&
  $19.24/0.15$ &$19.06/0.15$& $0.18/0.04$&26\nl

47 &3.58$^\circ$& 290$^\circ$ & 1.0114 & 141.7&
  $19.25/0.16$ &$19.07/0.13$& $0.19/0.04$&17\nl

80 &0.69$^\circ$& 308$^\circ$ & 0.9993 & 237.9&
  $19.37/0.19$ &$19.10/0.18$& $0.27/0.04$&54\nl
81 &0.70$^\circ$& 116$^\circ$ & 0.9991 & 210.6& 
  $19.39/0.28$ &$19.17/0.26$& $0.22/0.05$&38\nl

82 &0.58$^\circ$& 39$^\circ$ & 0.9931 & 170.4& 
  $19.38/0.18$ &$19.12/0.17$& $0.26/0.04$&37\nl
\bf ALL \rm &&&& & $19.35/0.20$& $19.13/0.19$& $0.22/0.05$& 663\nl

\enddata
\end{deluxetable}

\begin{deluxetable}{cccccccccc}
%\footnotesize
\tablecolumns{10}
\tablecaption{Fourier Parameters of M2 RR1 Variables}
%\tablewidth{Opt}
\tablehead{\colhead{Star} & 
 \colhead{Period} & \colhead{$A_{0}$} & \colhead{$A_{1}$} & \colhead{$R_{21}$} 
& \colhead{$R_{31}$} & \colhead{$R_{41}$} &\colhead{$\phi_{21}$} 
&\colhead{$\phi_{31}$} &\colhead{$\phi_{41}$} \\
\colhead{} & \colhead{N} & \colhead{$A_{v}$} & \colhead{$\sigma_{fit}$} & 
\multicolumn{6}{l}{----------------------------- $\sigma$ --------------------------------} 
 \\
\colhead{(1)}& \colhead{(2)}& \colhead{(3)}& \colhead{(4)}& \colhead{(5)}&
\colhead{(6)}& \colhead{(7)}& \colhead{(8)}& \colhead{(9)}& \colhead{(10)}
}
\startdata

V19          &0.319416&16.06&0.226&0.1624&0.0772&0.0507&5.21&3.74&2.40\nl
   & 114      & 0.441 &0.025 &0.0153&0.0164&0.0168&0.10&0.21&0.31\nl
V24          &0.358162&16.00&0.211&0.1010&0.0573&0.0245&4.58&4.06&3.09\nl
(LC450)   & 125      & 0.421 &0.031 &0.0200&0.0193&0.0195&0.19&0.34&0.77\nl

V32          &0.361938&16.05&0.212&0.0714&0.0845&0.0403&4.77&4.41&1.93\nl
(LC864)   & 145      & 0.429 &0.050 &0.0285&0.0290&0.0296&0.41&0.36&0.70\nl
\enddata
\tablecomments{LC99 discovered 13 new RR Lyrae variables in M2. In the
catalog of Variable Stars in Globular Clusters at 
http://www.astro.utoronto.ca/people.html (Clement et al. 2001), these
have been numbered V22 through V34.}
\end{deluxetable}

\begin{deluxetable}{lcccccccc}
%\footnotesize
\tablecolumns{9}
\tablecaption{Corrected Magnitudes and Reddenings by Field}
%\tablewidth{Opt}
\tablehead{\colhead{} & \multicolumn{4}{l}{--------------------All Stars--------------------} &
\multicolumn{4}{l}{-----------------M5-Like Stars-----------------} \\
\colhead{} & \colhead{} & 
\multicolumn{3}{c}{-----------------Mean-----------------} 
& \colhead{} & 
\multicolumn{3}{c}{-----------------Mean-----------------} \\
\colhead{Field} & \colhead{N} & \colhead{$<V>_F$} 
& \colhead{$E(V-R)$} 
& \colhead{$<V_0>_F$} & \colhead{N} & 
\colhead{$<V>_F$} & \colhead{$E(V-R)$} & \colhead{$<V_0>_F$}  \\
\colhead{(1)}& \colhead{(2)}& \colhead{(3)}& \colhead{(4)} &\colhead{(5)}
&\colhead{(6)}&\colhead{(7)}&\colhead{(8)}&\colhead{(9)}
}

\startdata
2  & 32 & 19.25$\pm0.03$ & 0.06$\pm0.01$ &18.92$\pm0.04$&  8&
19.23$\pm0.07$ & 0.06$\pm0.02$& 18.91$\pm0.08$\nl
3  & 22 & 19.50$\pm0.05$ & 0.13$\pm0.01$ &18.84$\pm0.05$&  4&
19.48$\pm0.05$ & 0.13$\pm0.02$& 18.80$\pm0.05$\nl
5  & 34 & 19.25$\pm0.03$ & 0.06$\pm0.00$ &18.92$\pm0.03$&  8&
19.23$\pm0.04$ & 0.08$\pm0.01$& 18.81$\pm0.05$\nl
6  & 36 & 19.36$\pm0.02$ & 0.09$\pm0.01$ &18.87$\pm0.03$&  9&
19.45$\pm0.05$ & 0.11$\pm0.01$& 18.86$\pm0.07$\nl
10 & 14 & 19.22$\pm0.04$ & 0.06$\pm0.01$ &18.93$\pm0.03$&  6&
19.23$\pm0.07$ & 0.05$\pm0.01$& 18.94$\pm0.06$\nl
11 & 25 & 19.36$\pm0.05$ & 0.05$\pm0.01$ &19.10$\pm0.06$&  2& 
19.27$\pm0.08$ & 0.06$\pm0.05$& 18.94$\pm0.20$\nl
12 & 13 & 19.41$\pm0.04$ & 0.09$\pm0.01$ &18.94$\pm0.04$&  2&
19.42$\pm0.15$ & 0.10$\pm0.03$& 18.90$\pm0.02$\nl
13 & 32 & 19.31$\pm0.02$ & 0.08$\pm0.01$ &18.91$\pm0.03$&  8&
19.34$\pm0.05$ & 0.08$\pm0.01$& 18.92$\pm0.06$\nl
14 & 13 & 19.30$\pm0.04$ & 0.03$\pm0.01$ &19.15$\pm0.05$&  5&
19.26$\pm0.05$ & 0.04$\pm0.01$& 19.05$\pm0.07$\nl
15 & 13 & 19.56$\pm0.04$ & 0.11$\pm0.01$ &19.00$\pm0.03$&  3&
19.62$\pm0.09$ & 0.11$\pm0.01$& 19.02$\pm0.02$\nl
18 & 11 & 19.27$\pm0.04$ & 0.06$\pm0.01$ &18.97$\pm0.09$&  2&
19.35$\pm0.05$ & 0.05$\pm0.00$& 19.07$\pm0.05$\nl
19 & 13 & 19.22$\pm0.04$ & 0.03$\pm0.01$ &19.05$\pm0.05$&  4&
19.33$\pm0.03$ & 0.03$\pm0.01$& 19.14$\pm0.07$\nl
47 & 12 & 19.25$\pm0.05$ & 0.05$\pm0.01$ &18.99$\pm0.04$&  2&
19.28$\pm0.15$ & 0.03$\pm0.03$& 19.11$\pm0.04$\nl
80 & 33 & 19.39$\pm0.04$ & 0.13$\pm0.01$ &18.72$\pm0.04$& 10&
19.43$\pm0.06$ & 0.12$\pm0.02$& 18.79$\pm0.05$\nl
81 & 14 & 19.45$\pm0.05$ & 0.07$\pm0.01$ &19.05$\pm0.07$&  3&
19.50$\pm0.13$ & 0.09$\pm0.02$& 19.00$\pm0.23$\nl
82 & 13 & 19.37$\pm0.03$ & 0.11$\pm0.01$ &18.76$\pm0.06$&  4&
19.43$\pm0.06$ & 0.13$\pm0.01$& 18.73$\pm0.11$\nl
{\bf ALL} & 330 & 19.34$\pm0.01$ & 0.08$\pm0.003$&18.92$\pm0.01$ & 80 &
19.35$\pm0.02$ & 0.09$\pm0.005$& 18.91$\pm0.02$ \nl 

\enddata

\tablecomments{In each case, the errors listed represent the standard
error of the mean. However,
for both the extinction and $V_0$, there are systematic errors as well: $0.022$
mag in $E(V-R)$, which propagates to an error of $0.12$ mag in $V_0$}
\end{deluxetable}

\begin{deluxetable}{ccccccc}
%\footnotesize
\tablecolumns{7}
\tablecaption{Mean Magnitudes and Colors for the M5-like RR1 Stars in
Field 80}
%\tablewidth{Opt}
\tablehead{\colhead{Star} & 
 \colhead{$<V>_F$} & \colhead{$<V>_{int}$} & \colhead{$<R>_F$} & 
 \colhead{$<R>_{int}$} & \colhead{$<V>_F-<R>_F$} & \colhead{$<V-R>_{mag}$} 
 \\
\colhead{(1)}& \colhead{(2)}& \colhead{(3)}& \colhead{(4)}& \colhead{(5)}&
\colhead{(6)}& \colhead{(7)}
}
\startdata

80.6351.2358 & 19.32 & 19.31 & 19.10 & 19.09 & 0.22 & 0.22 \nl
80.6354.3658 & 19.81 & 19.79 & 19.48 & 19.48 & 0.33 & 0.32 \nl
80.6475.3548 & 19.46 & 19.45 & 19.20 & 19.20 & 0.26 & 0.25 \nl
80.6589.2425 & 19.56 & 19.54 & 19.30 & 19.29 & 0.26 & 0.26 \nl
80.6596.3127 & 19.51 & 19.50 & 19.24 & 19.23 & 0.27 & 0.27 \nl
80.6710.2075 & 19.41 & 19.40 & 19.18 & 19.17 & 0.23 & 0.23 \nl
80.6832.2030 & 19.59 & 19.58 & 19.31 & 19.30 & 0.28 & 0.27 \nl
80.7192.4927 & 19.49 & 19.48 & 19.19 & 19.18 & 0.30 & 0.30 \nl
80.7320.1224 & 19.26 & 19.25 & 19.03 & 19.02 & 0.23 & 0.23 \nl
80.7437.1665 & 19.39 & 19.38 & 19.11 & 19.10 & 0.29 & 0.29 \nl

\enddata
\end{deluxetable}

\begin{deluxetable}{lcccccc}
%\footnotesize
\tablecolumns{7}
\tablecaption{Mean Magnitudes for Stars Observed by C03}
%\tablewidth{Opt}
\tablehead{\multicolumn{3}{c}{---------------------MACHO----------------------} 
& \multicolumn{3}{c}{--------------C03--------------} & \\
\colhead{ID} & \colhead{$<V>$} & \colhead{$N_{obs}$}
& \colhead{ID} & \colhead{$<V>$} & \colhead{$N_{obs}$} 
& \colhead{$\Delta <V>$}\\
\colhead{(1)}& \colhead{(2)}& \colhead{(3)} & \colhead{(4)} & \colhead{(5)}
& \colhead{(6)} & \colhead {(7)}
}
\startdata

6.6689.563*  & 19.22 & 443 & 2249 & 19.39 & 70 & 0.17\dag \nl
13.6689.3055* & 19.24 & 393 &     &       &    & 0.15\dag \nl
6.6812.1063 & 19.61 & 240 & 8837 & 19.58 & 64 & -0.03 \nl
6.7054.710*  & 19.43 & 318 & 7864 & 19.48 & 67 & 0.05 \nl
13.7054.3006* & 19.36 & 171 &     &       &    & 0.12 \nl
13.5838.667 & 19.39 & 253 & 7648 & 19.40 & 69 &  0.01 \nl
13.5959.584 & 19.20 & 423 & 7783 & 19.30 & 70 & 0.10 \nl
13.6079,604 & 19.25 & 373 & 4749 & 19.32 & 63 & 0.07 \nl
13.6201.670 & 19.10 & 462 & 7490 & 19.18 & 70 & 0.08 \nl
\bf Mean \rm (all) &      &      &     &&  & 0.07 \nl
\bf Mean \rm (ex 2249) &      &   &&   &       & 0.05 \nl
\enddata

\tablecomments{*MACHO 6.6689.563 is the same star as 13.6689.3055 and
6.7054.710 is the same star as 13.7054.3006.
\dag C03 found excessive scatter in the light curve for their
star \#2249.}

\end{deluxetable}

\clearpage

\centerline{\bf{FIGURES}}

\begin{figure}
\epsscale{0.8}
\plotone{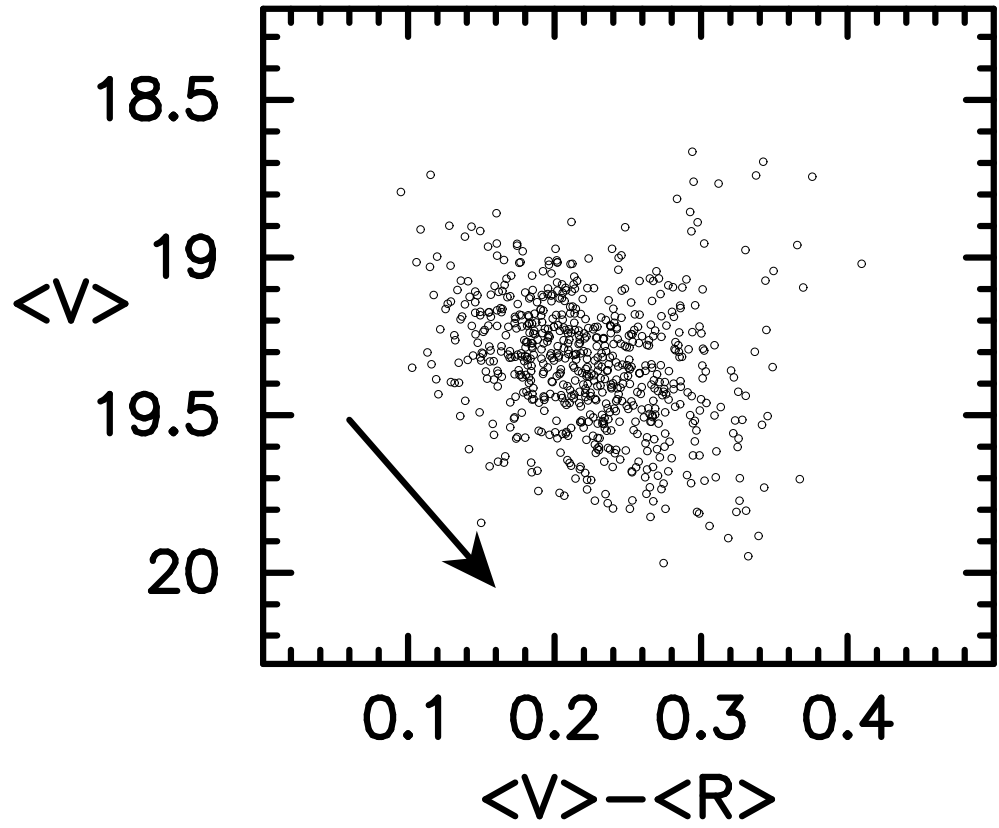}
\caption{$<V>_F$ vs $<V>_F-<R>_F$ for the stars listed in Table 1. The reddening vector
($A_V/E(V-R)=5.35$) is marked with an arrow.}
\end{figure}

\begin{figure}
\epsscale{0.6}
\plotone{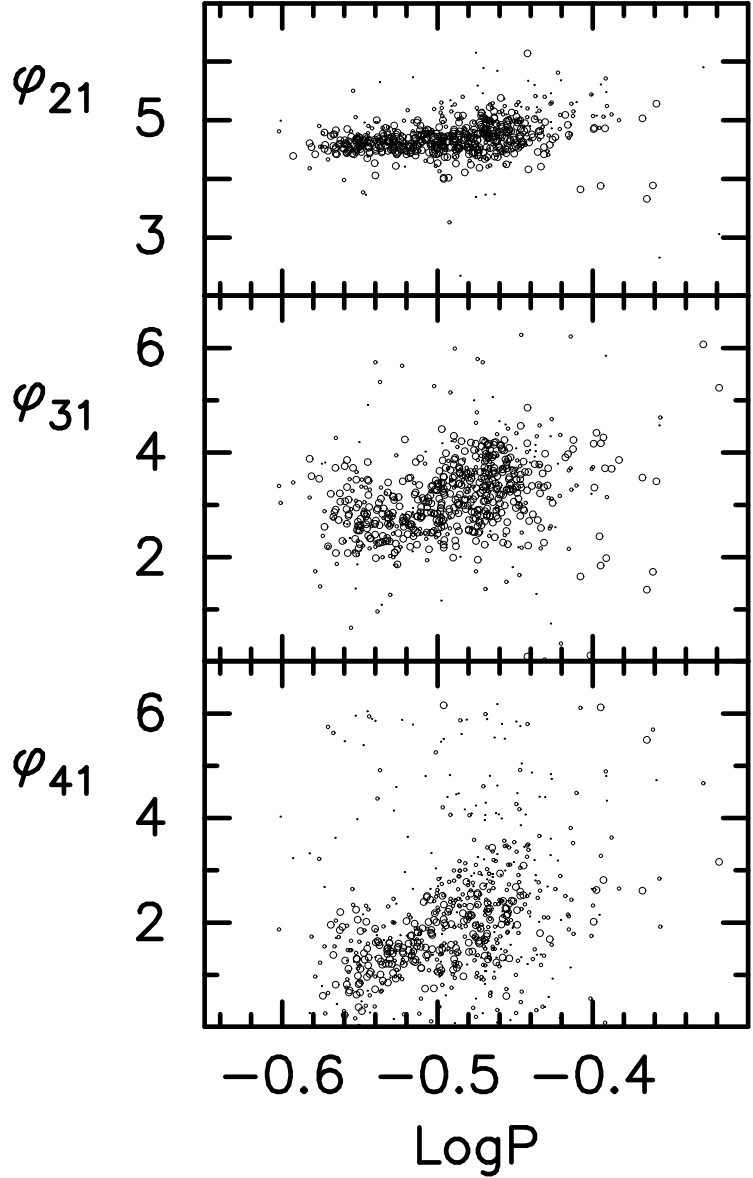}
\caption{Fourier phase differences $\phi_{21}$, $\phi_{31}$ and $\phi_{41}$  as a 
function of $\log P$ for the $V$ data of the 663 program stars summarized
in Table 8. 
The plotted points are open
circles that have 3 sizes: the larger the size, the lower the error. 
For $\phi_{21}$, the largest size denotes
standard errors less than $0.2$, the smallest size denotes
errors greater than $0.4$ and the intermediate size denotes
errors between $0.2$ and $0.4$.
For the $\phi_{31}$ and $\phi_{41}$ plots, the three sizes
denote standard errors less than $0.4$, 
greater than $0.8$ and between $0.4$ and $0.8$, respectively.}
\end{figure}

\begin{figure}
\epsscale{0.7}
\plotone{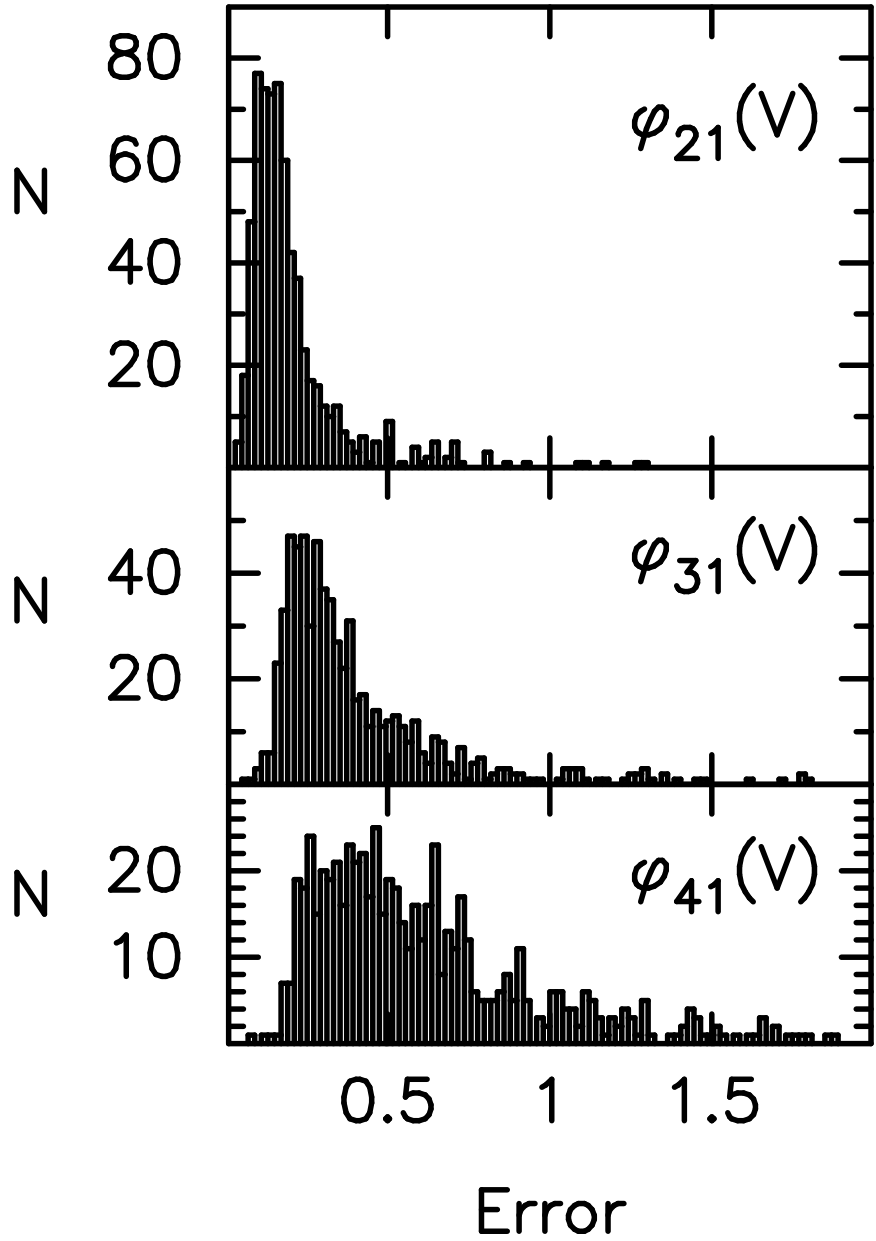}
\caption{Histograms illustrating the distribution of the standard errors in 
$\phi_{21}$, 
$\phi_{31}$ and $\phi_{41}$ for the $V$ data that are plotted in Fig. 2. }
\end{figure}

\begin{figure}
\epsscale{0.6}
\plotone{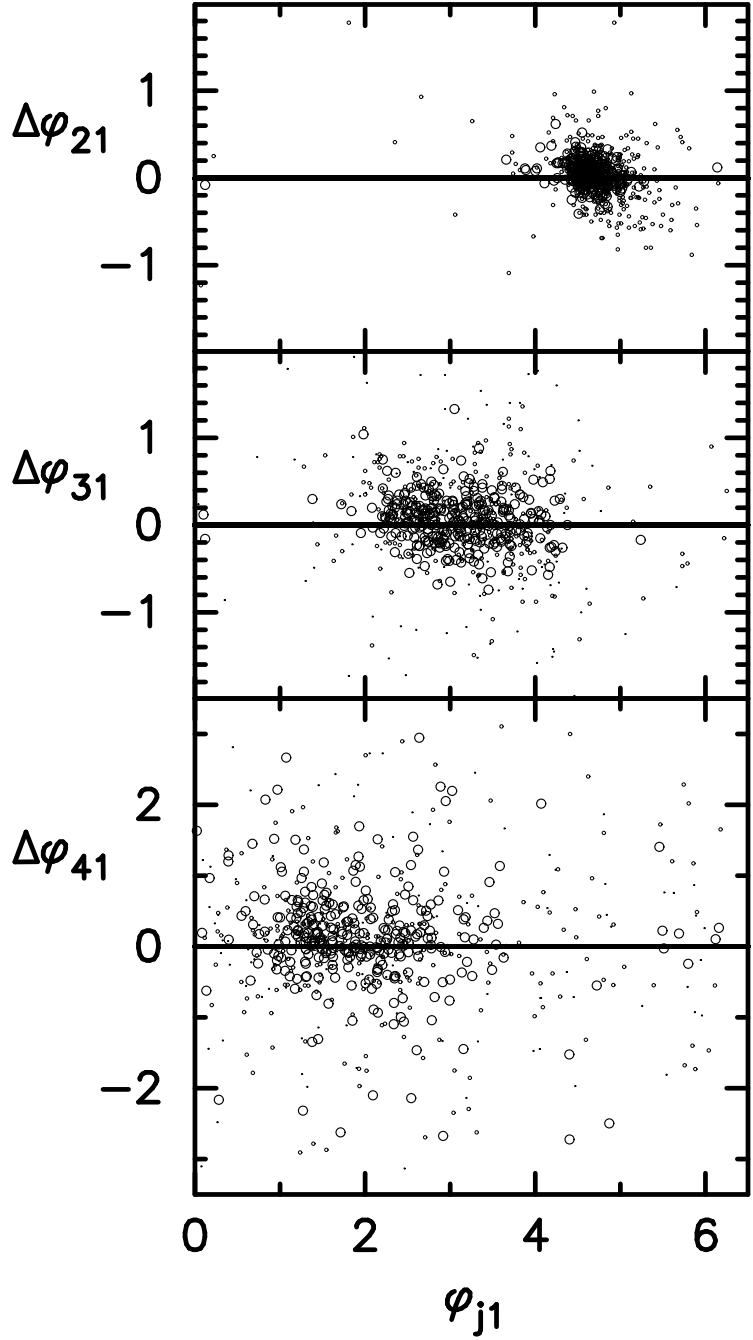}
\caption{Plots of $\Delta \phi_{21}$ vs $\phi_{21}$(V), 
$\Delta \phi_{31}$ vs $\phi_{31}$(V)
and $\Delta \phi_{41})$ vs $\phi_{41}$(V), 
to illustrate the
effect of color on the Fourier phase differences. 
The $\Delta \phi_{j1}$ values denote [$\phi_{j1}(R)-\phi_{j1}(V)]$. 
The horizontal lines
are drawn at $\Delta \phi_{j1} =0$.
The scaling of the plotted points (open circles)
is the same as in Fig. 2, i.e. the different sizes
correspond to the errors in the Fourier coefficients for the $V$ data.}
\end{figure}

\begin{figure}
\epsscale{0.5}
\plotone{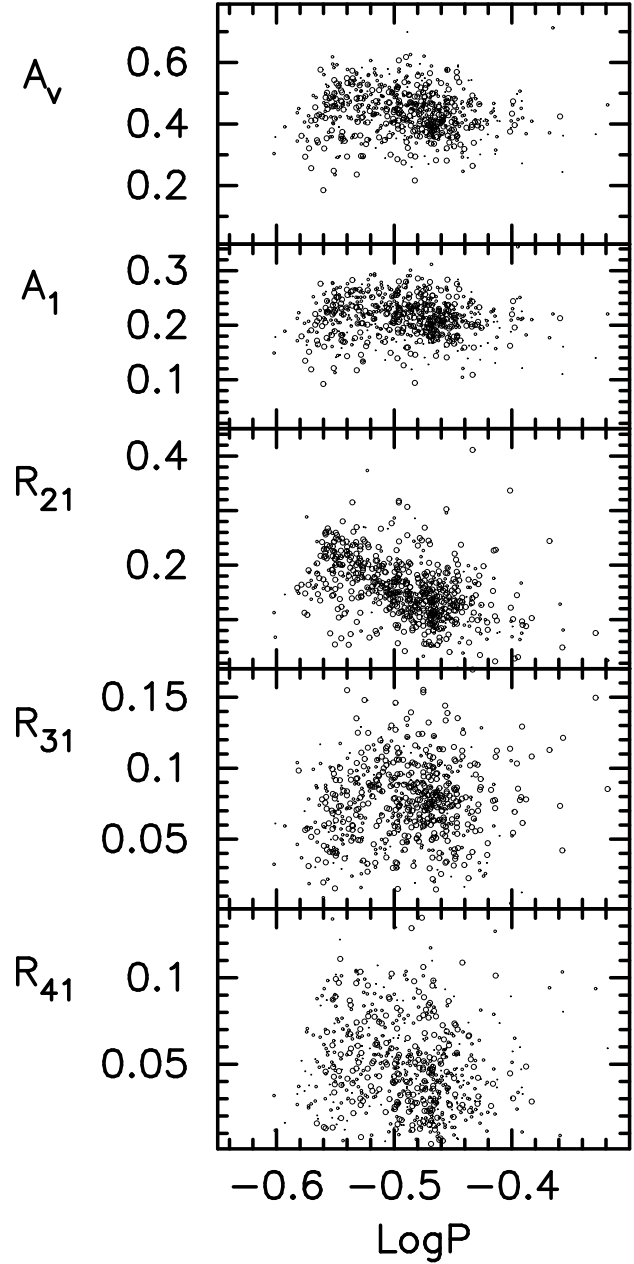}
\caption{V amplitude, Fourier amplitude $A_1$ and amplitude ratios $R_{21}$, $R_{31}$
and $R_{41}$ for the $V$ data as a function of $\log P$. 
The plotted points are open circles that have three sizes,
as in Figs. 2 and 4. 
For $A_V$ and $A_1$, the largest size denotes stars for which the standard
deviation ($\sigma$) of the fit to equation (1) is less than $0.06$, 
the smallest size denotes $\sigma$ greater than $0.08$ and the intermediate
size denotes $\sigma$ between $0.06$ and $0.08$.
For the $R_{21}$, $R_{31}$ and $R_{41}$ plots, the scale of the
points is the same as in Fig. 2. It relates to the errors in the Fourier phase
differences for the $V$ data.}
\end{figure}

\begin{figure}
\epsscale{0.7}
\plotone{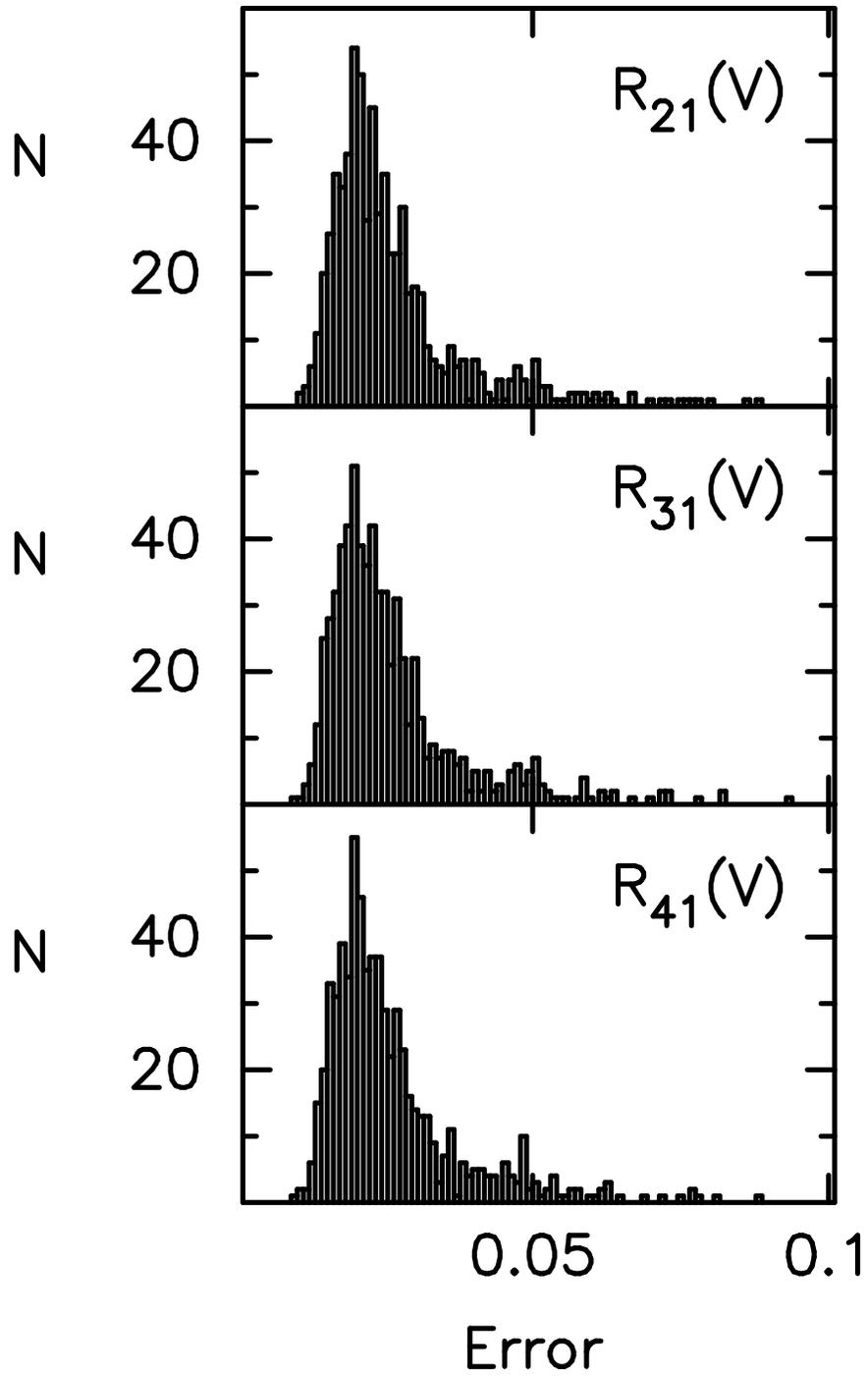}
\caption{Histograms illustrating the distribution of the standard
  errors in $R_{21}$, $R_{31}$ and $R_{41}$ for the $V$ data plotted in Fig. 5.}\end{figure}

\clearpage

\begin{figure}
\epsscale{0.65}
\plotone{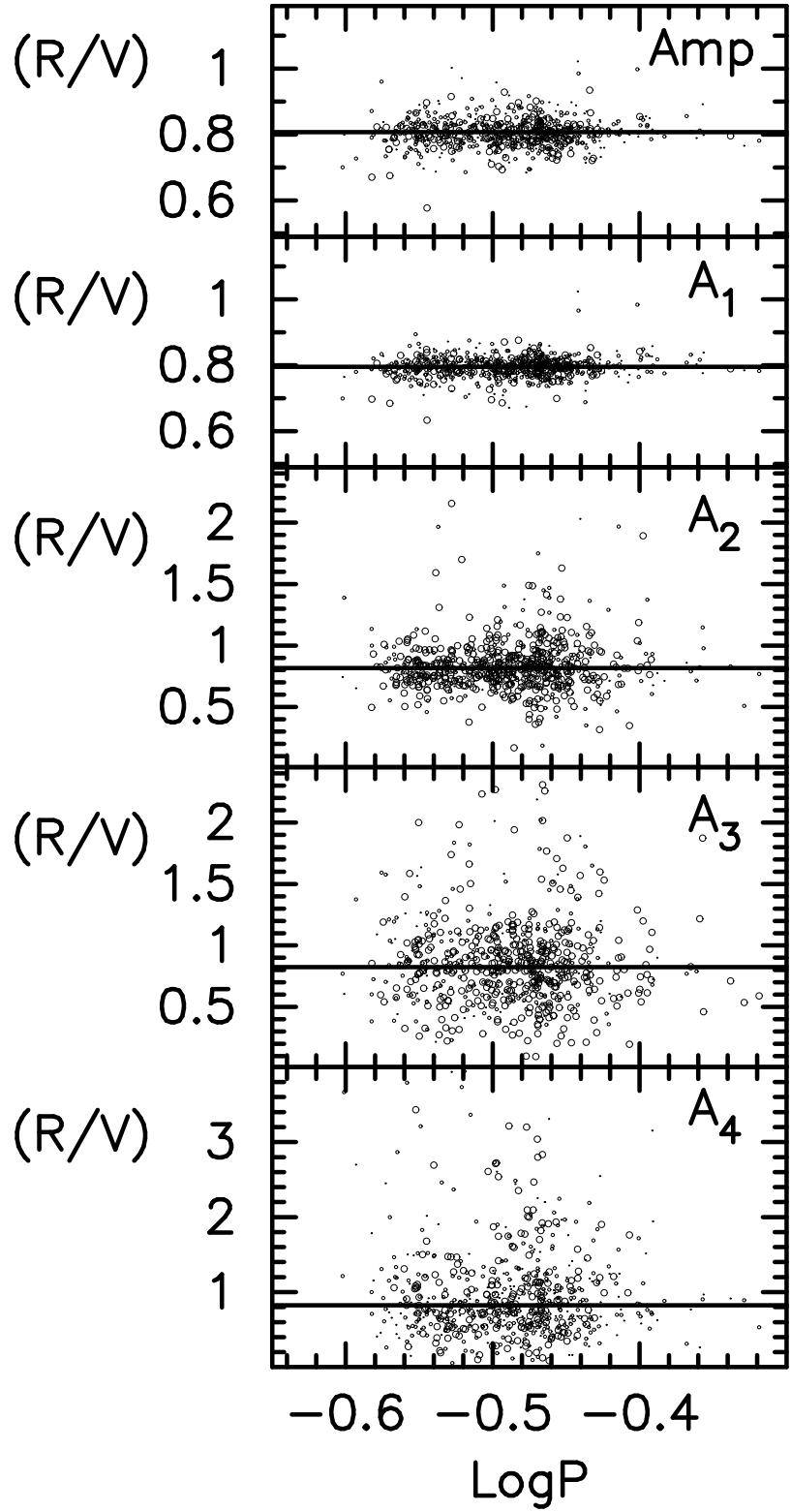}
\caption{Plots of the ratios: $A_R/A_V$, $A_1 (R)/A_1 (V)$, $A_2(R)/ A_2 (V)$,  $A_3 (R)/ A_3 (V)$ and $A_4 (R)/ A_4 (V)$ to show the effect 
of wavelength band. The symbols are the same as in Figure 5.}
\end{figure}

\begin{figure}
\epsscale{0.4}
\plotone{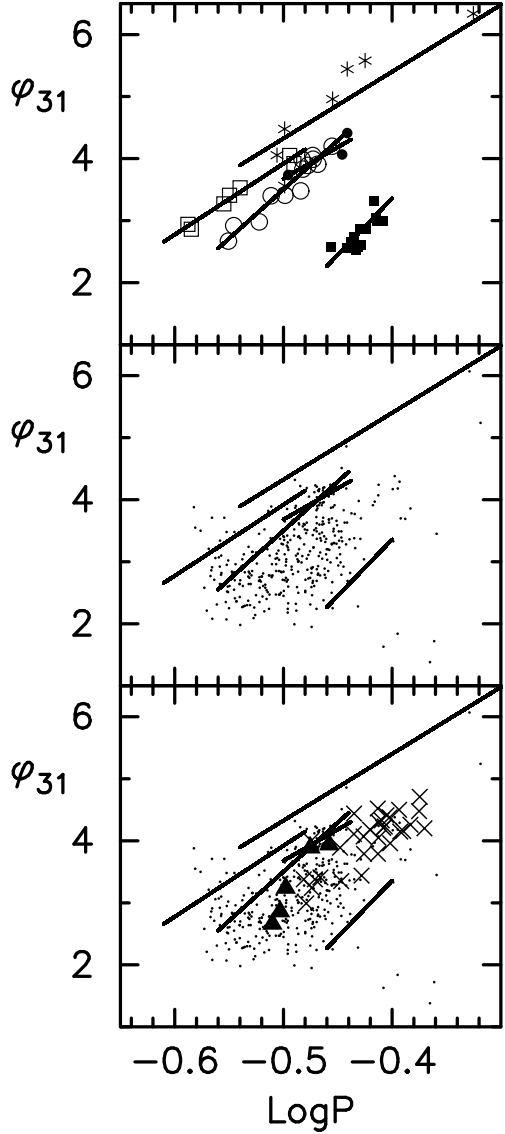}
\caption{In the upper panel, we plot $\phi_{31}$ vs $\log P$ for the RR1 variables in
five well studied Galactic globular clusters. The
different symbols represent different clusters: asterisks for NGC 6441, open
squares for M107 (NGC 6171), open circles for M5, solid circles for M2
and solid squares for M68.  For each cluster, a straight 
line derived from a least squares fit is plotted through the points. 
In the central panel, these lines are repeated and the
points for the RR1 variables in the LMC for which the error in $\phi_{31}$
is less than 0.4 are also plotted. The lower panel is the same as
the central panel with the RR1 stars in two additional clusters, $\omega$ 
Centauri (crosses) and M3 (solid triangles) included.}
\end{figure}

\begin{figure}
\epsscale{0.5}
\plotone{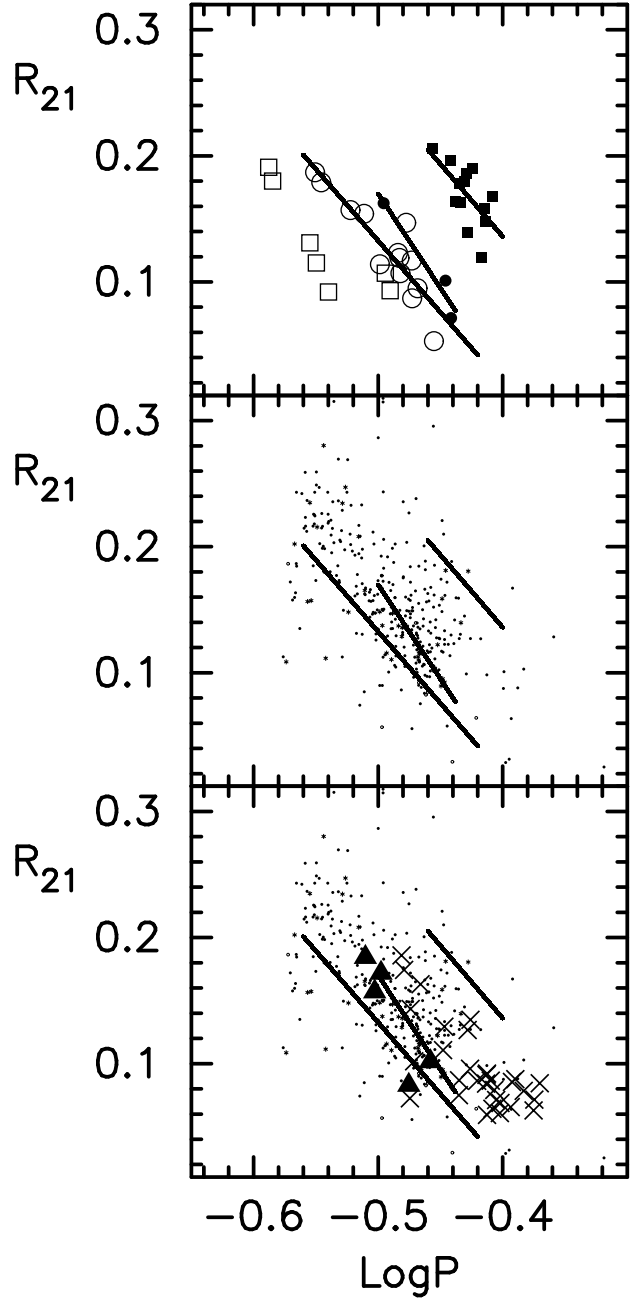}
\caption{In the upper panel, we plot $R_{21}$ vs $\log P$ for the RR1 variables in
four clusters: open
squares for M107 (NGC 6171), open circles for M5, solid circles for M2 
and solid squares for M68.  For M5, M2 and M68, a straight 
line derived from a least squares fit is plotted through the points. 
In the central panel, these lines are repeated and the
points for the RR1 variables in the LMC for which the error in $R_{21}$
is less than $0.025$ are also plotted. The lower panel is the same as
the central panel with the RR1 stars in two additional clusters, $\omega$ 
Centauri (crosses) and M3 (solid triangles) included. }
\end{figure}

\begin{figure}
\epsscale{0.8}
\plotone{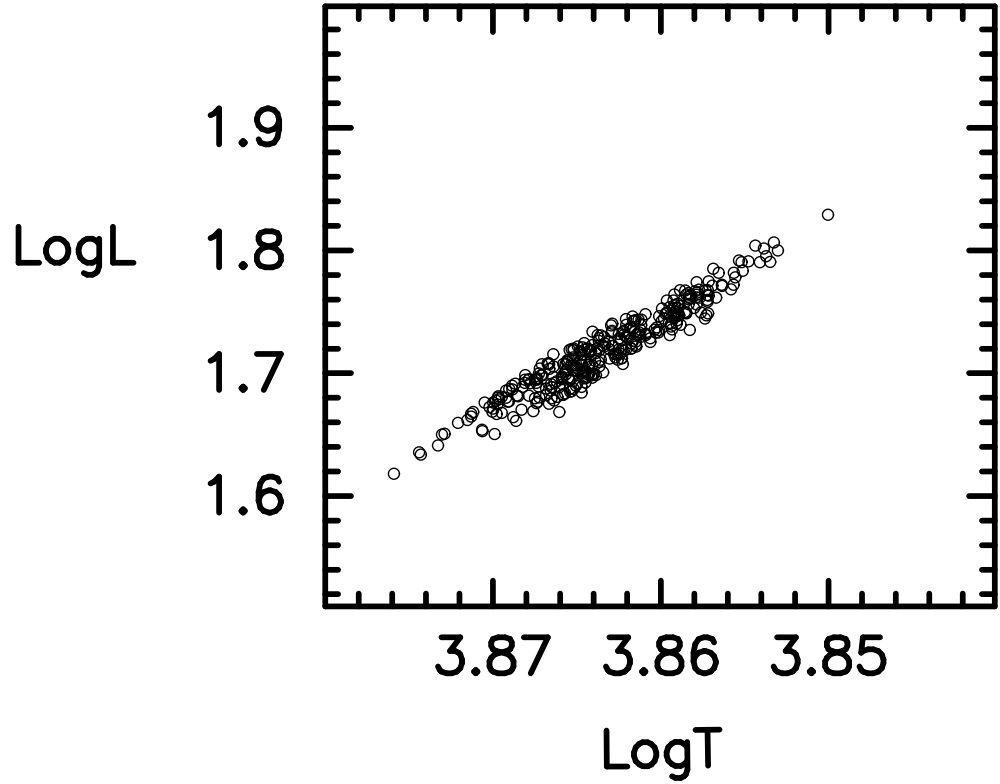}
\caption{$\log L/\lsun$ calculated from equation (2) versus $\log T_{eff}$ 
calculated from equation (5) for the $V$ data.  Only
the 330 stars for which the period is in the range $-0.56 < \log P < -0.4$,
$\sigma (\phi_{31})<0.4$, 
the amplitude $A_V >0.3$ and the amplitude ratio $A_R/A_V$ is greater than
$0.75$ and less than $0.85$ are included. These are the stars considered
to be RR1 variables and are referred to as the bona fide RR1 variables 
throughout the  paper.}
\end{figure}

\begin{figure}
\epsscale{1.0}
\plotone{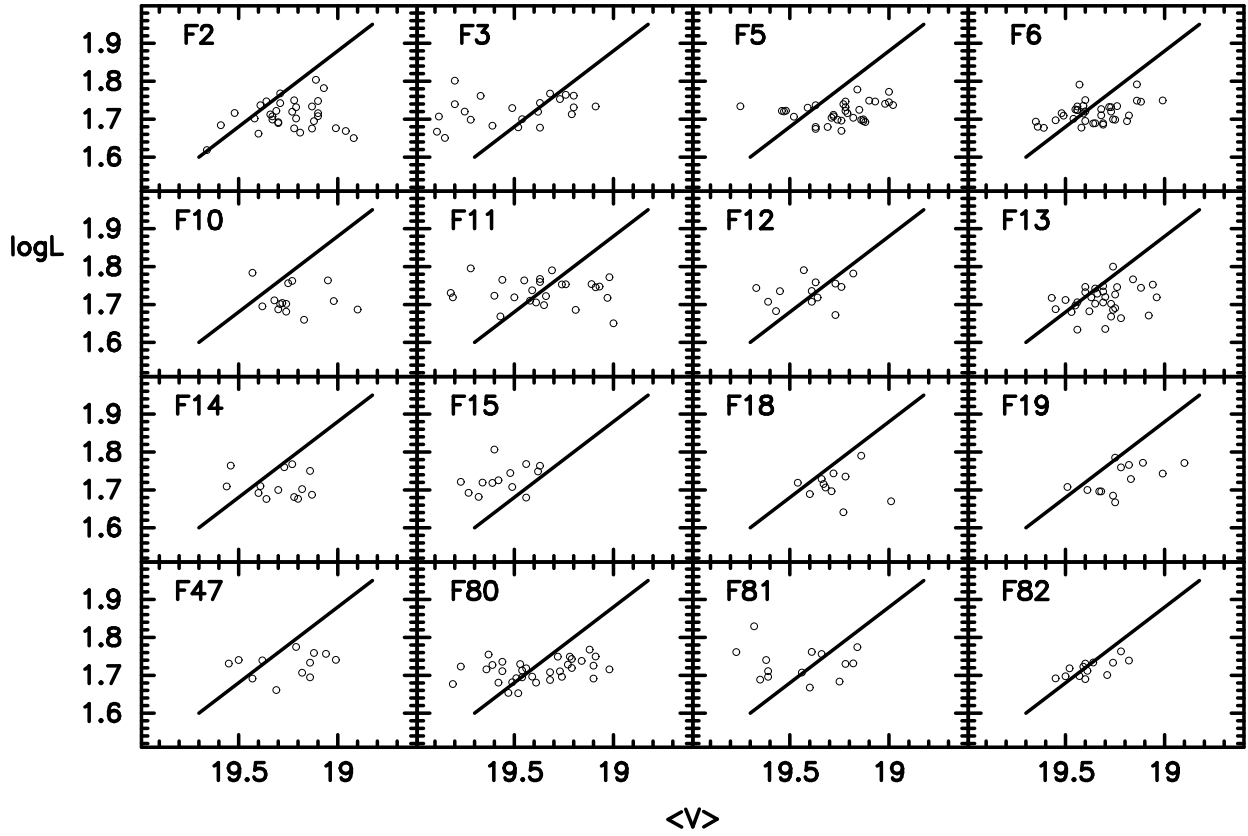}
\caption{Plots of $\log L/\lsun$ versus
the mean $V$ magnitude for the 330 RR1 variables plotted in Figure 10.
The stars in each field are plotted separately and the 
line drawn through the points in each panel has a slope 
$\Delta \log L/\Delta V=-0.4$. The lines are plotted at the same position
in each panel so that differences in $<V>$ among the fields can
be readily observed.}
\end{figure}

\begin{figure}
\epsscale{1.0}
\plotone{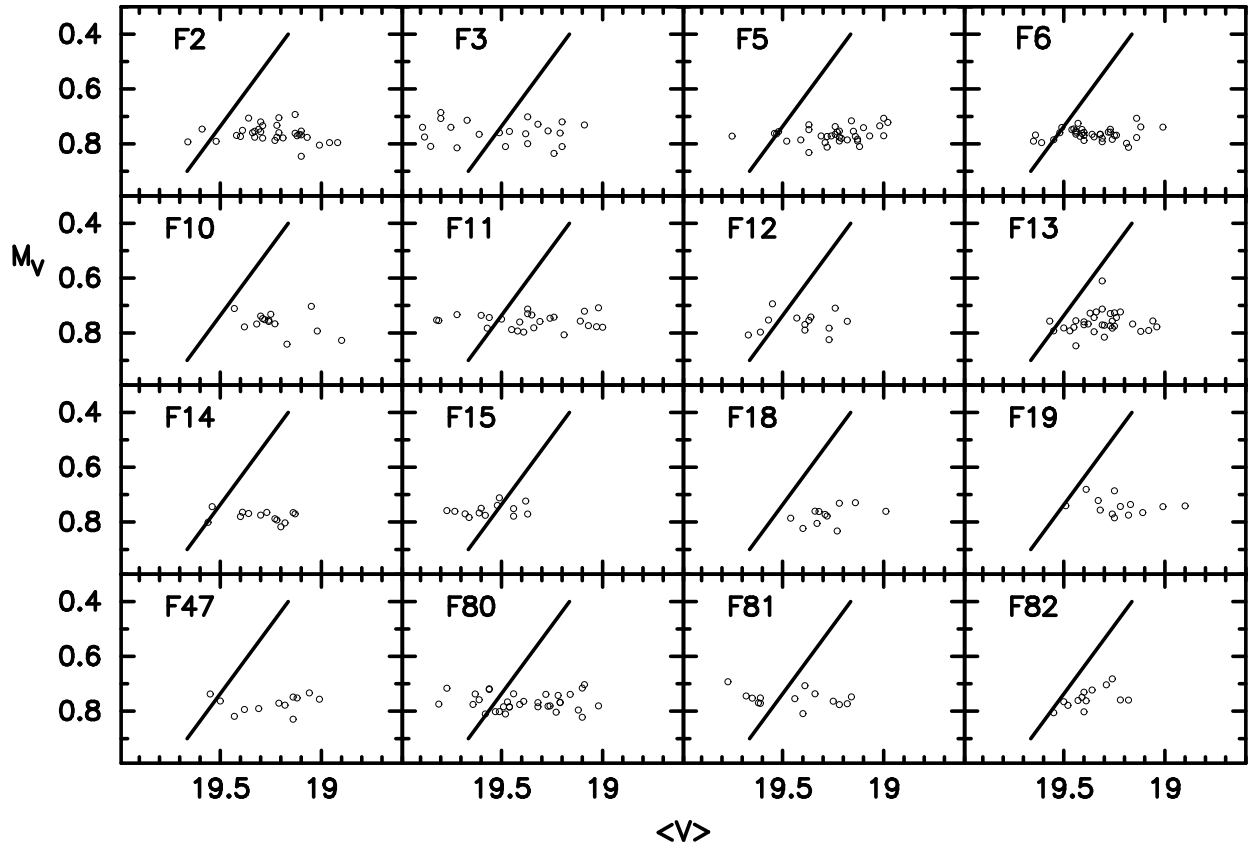}
\caption{Plots of $M_V$ calculated from equation (3) versus $<V>$
for the 330 RR1 variables plotted in Figure 10. 
The stars in each field are plotted separately and the line drawn
through the points in each panel has a slope 
$\Delta M_V /\Delta V=1$.}
\end{figure}

\begin{figure}
\epsscale{0.8}
\plotone{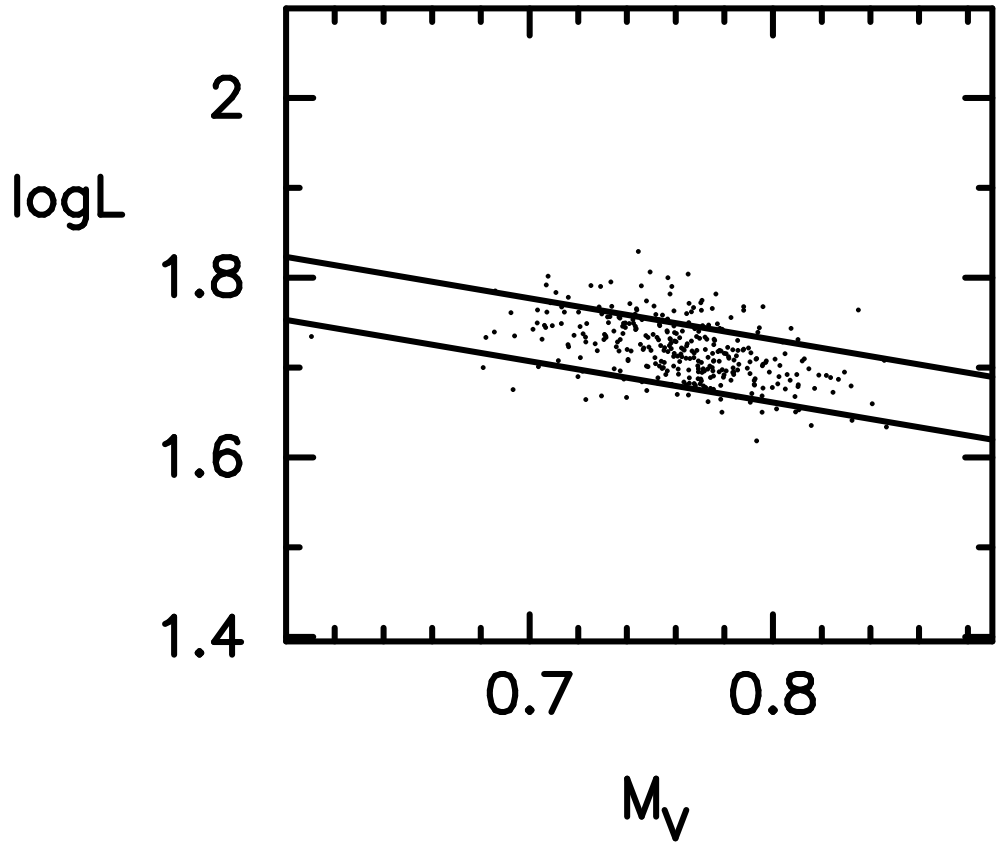}
\caption{$\log L/\lsun$ versus $M_V$ calculated from equations (2) and
(3) using the $V$ data for the 330 RR1 variables plotted in Figure 10.
The envelope lines have a slope of $-0.46$, the predicted slope for
$\Delta \log L/ \Delta M_V$ and
are separated by $\Delta \log L = 0.07$.}
\end{figure}

\begin{figure}
\epsscale{1.0}
\plotone{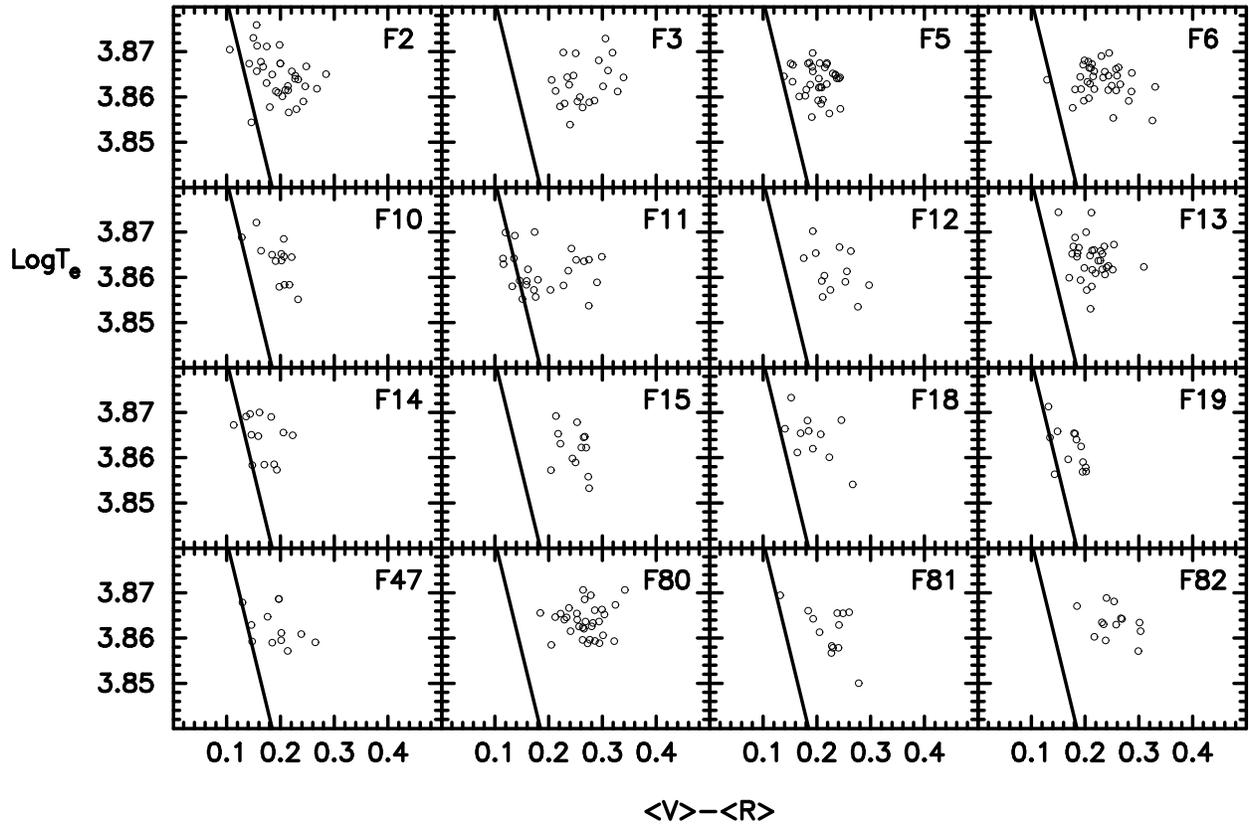}
\caption{$\log T_{eff}$ versus [$<V>_F-<R>_F$] 
for the 330 RR1 variables plotted in Figure 10. 
The stars in each field are plotted separately to show the differences
in color excess. The ridge lines
represent the $\log T_{eff}-(V-R)_0$ relations derived from
equation (6) assuming $\log g=2.9$ and [M/H]$=-1.5$.}
\end{figure}

\begin{figure}
\epsscale{1.0}
\plotone{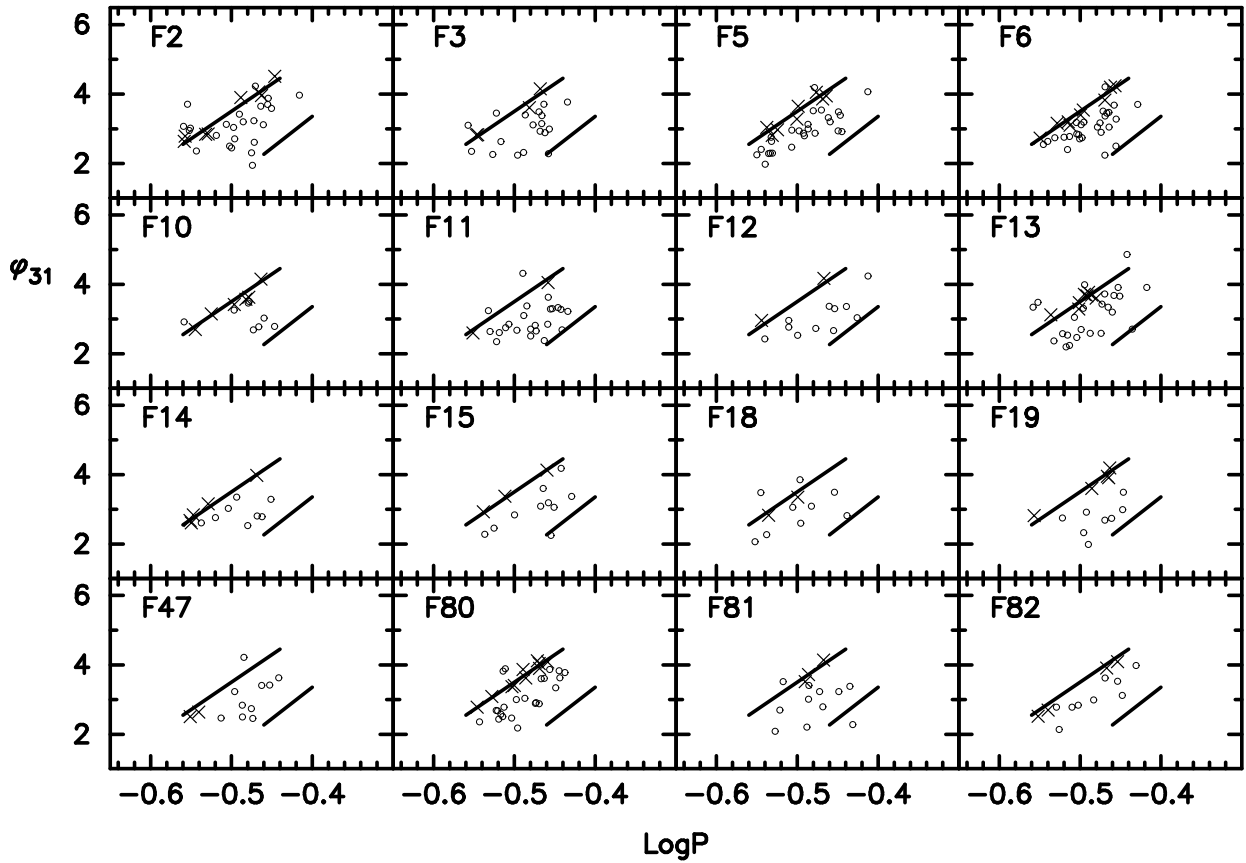}
\caption{Plots of $\phi_{31}$ vs $\log P$ for the 330 RR1 variables 
plotted in Figure 10, with each field plotted separately. 
The straight lines are the M5 and M68 lines from
Fig. 8. The `M5-like' variables are designated as crosses.}
\end{figure}

\begin{figure}
\epsscale{0.25}
\plotone{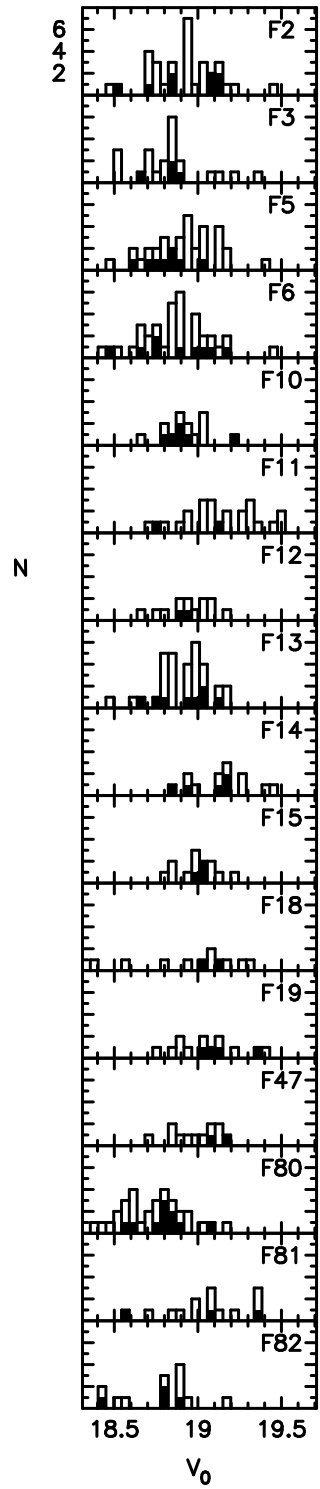}
\caption{The distribution of $V_0$ for the 330 RR1 variables plotted in
Figure 10, with each field
plotted separately. The solid areas represent the `M5-like' variables.}
\end{figure}

\begin{figure}
\epsscale{0.8}
\plotone{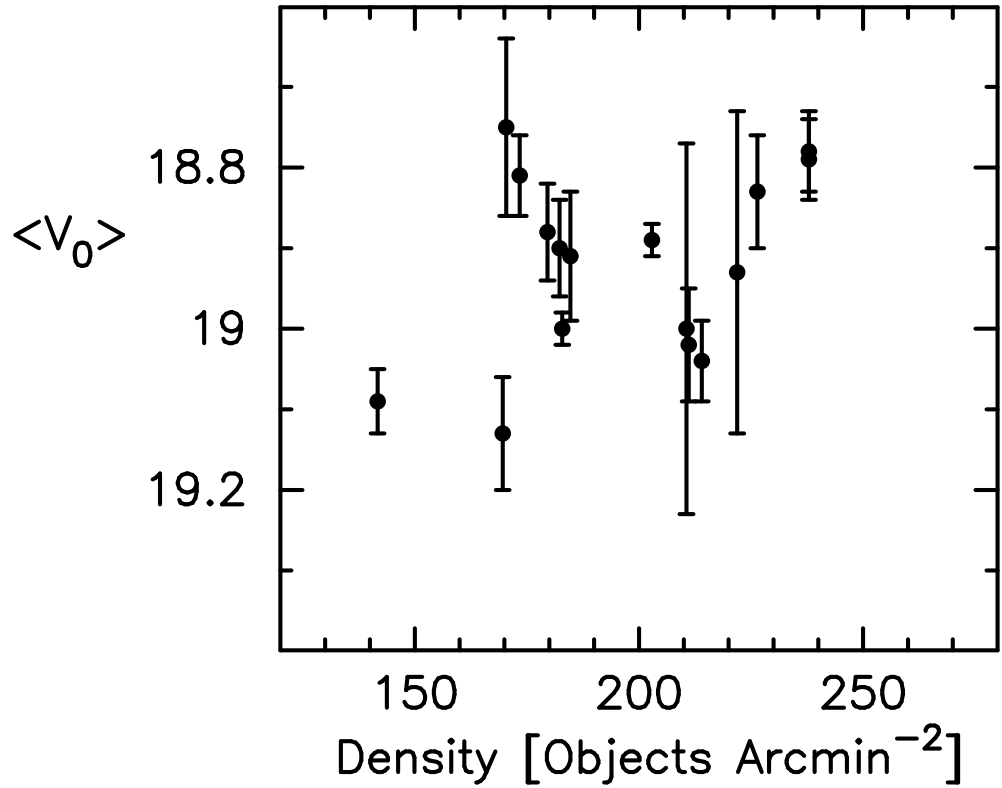}
\caption {The mean $V_0$ for the M5-like variables 
versus the density for the 16 fields. 
The mean $V_0$ value for each field has
been adjusted to compensate for the distance $D/D_0$, listed in column
(4) of Table 8 and the error bar represents the standard 
deviation of $V_0$ within the field.
The density is the average number of objects per square arcmin, listed in
column (5) of Table 8. }
\end{figure}

\end{document}